\documentclass[aps, prb, twocolumn]{revtex4-2}
\usepackage{bm}
\usepackage[toc,page]{appendix}
\usepackage{comment}
\usepackage{dcolumn}
\usepackage{color,hyperref}
\usepackage[caption=false]{subfig}
\hypersetup{colorlinks=true, linkcolor=blue, citecolor=blue, urlcolor=blue} 
\usepackage{amsfonts}
\usepackage{amsmath}
\usepackage{graphicx}
\usepackage{dcolumn}
\usepackage{bm}
\usepackage{physics}
\usepackage{braket}
\usepackage{xcolor}
\usepackage{soul}

\begin{document}
\title{Topological initial state independence in Boundary Time Crystals}
\author{Dominik Nemeth}
\email{dominik.nemeth@manchester.ac.uk}
\author{Ahsan Nazir}
\author{Alessandro Principi}
\author{Robert-Jan Slager}
\affiliation{Department of Physics and Astronomy, The University of Manchester, Oxford Road, Manchester M13 9PL, United Kingdom}

\date{\today}

\begin{abstract}
Boundary time crystals (BTCs) break time-translation symmetry and exhibit long-lived, robust oscillations insensitive to initial conditions. We show that collective spin BTCs can admit emergent topological winding numbers in operator space. Expanding the density operator in a spherical tensor basis, we map the Lindblad dynamics onto an effective local hopping problem, where collective degrees of freedom label sites of an emergent operator space lattice. Whenever the emergent winding number is non-zero, topological obstructions prevent the localization of operator modes on the lattice and instead enforce their delocalization across operator space. This delocalization enables non-unital sourcing from the identity sector to propagate across operator space and continuously feed the observable sector in an initial state independent manner, thereby generating robust oscillatory responses in the time crystal phase. Our results frame BTC dynamics as a form of topologically constrained operator space transport and suggest a close connection to non-Hermitian skin effects.
\end{abstract}

\maketitle

\section{Introduction}
\label{sec:introduction}
Time crystals have emerged as a central concept in nonequilibrium quantum physics, revealing new forms of dynamical order characterized by persistent oscillations in time \cite{Wilczek2012,Bruno2013, Watanabe2015, Khemani2016, Khemani2019, Sacha2015, Sacha2018, Else2016, Else2017, Else2020, Keyserlingk2016, Yao2017, Lazarides2017, Zaletel2023, Zhang2017}. Beyond their original formulation in closed and periodically driven systems, increasing attention has focused on time-crystal phases in open quantum systems, where dissipation and decoherence can play a constructive role in realizing long-lived oscillations \cite{Seibold2020, Lledo2020, Booker2020, Passarelli2022, Carollo2022, Nakanishi2023, Nemethb2025}.

A particularly striking realization is the boundary time crystal (BTC) \cite{Iemini2018}, which arises in collective spin systems weakly-coupled to Markovian environments. In the mean-field limit of this model, oscillations persist indefinitely in the thermodynamic limit despite the presence of dissipation~\cite{Iemini2018, Piccito2021}, making BTCs robust realizations of time-crystalline behavior in many-body systems. Recent developments have shown that while mean-field approximations yield the exact description for initially uncorrelated states in the thermodynamic limit~\cite{Carollo2021, Carollo2022, Fiorelli2023}, they fail to capture the dynamics of correlated initial states~\cite{Piccito2021, Mukherjee2024}. In response to this, several works have gone beyond the mean-field limit~\cite{Lourenco2022, Nemethb2025, Liu2025}, in particular it was shown that a fully quantum description is possible via spherical tensor operators~\cite{Nemeth2026}.

On a seemingly distinct front, topological concepts have provided a powerful framework for understanding properties and responses of condensed matter systems by focusing on the global structure of eigenstates, rather than microscopic details. In terms of systems that admit an effective band description, a rather unified understanding has emerged~\cite{Kruthoff2017,Bradlyn2017, Bouhon2020b,Po2017, Brouwer2023} that has recently been related to the wider concept of quantum geometry ~\cite{Provost1980,Resta_2011}. These geometric interpretations are increasingly being proven to be valuable in the study of various types of effective responses~\cite{Torma2023,Bouhon2023,Ahn2020,Yu2025qgt}. This topological lens has additionally also been the subject of active interest in non-Hermitian (NH) systems, that is systems described by an effective NH Hamiltonian obtained by neglecting the Lindbladian quantum jump term~\cite{Hatano1996,Bender_2007,Kunst2021}. Due to the non-Hermiticity and resulting complex energies, usual line-gap classifications can in such systems be extended by point-gap topologies, in which a winding number around a singular point is defined~\cite{Zhou2019,Kawabata2019}. Such point-gap topologies can have a striking consequence in terms of inducing a NH skin effect~\cite{Borgnia2020,Okuma2020,Hatano1996}, whereby a net drift across a chain forces the accumulation of eigenmodes at the boundary of this chain.

Recent work has demonstrated that key concepts of NH topology, such as winding numbers and the skin effect, can be extended to Lindbladian superoperators \cite{Prosen2008, Prosen2010, Diehl2011, Bardyn2013,Lieu2020,Minganti2020,Michishita2020, Wanjura2020,Haga2021,Zhou2022,Yang2022, Chaduteau2026}. In the class of quadratic Lindbladians, a direct relation can be established between the winding numbers of the Lindbladian and those of the associated postselected NH Hamiltonian \cite{Chaduteau2026}. While these quadratic systems possess an intrinsic notion of spatial locality, an analogous concept can emerge in collective spin systems, where locality is instead realized in operator space \cite{Nemeth2026}.

In this work, we uncover that collective spin BTC dynamics can be understood in terms of \emph{operator space topology}. It is well known that collective spin dynamics can be recast as effective hopping problems in a suitable basis~\cite{Wood2014, Ribeiro2019}, and more recently extended to operator space in BTCs~\cite{Nemeth2026}. In this representation, dissipative processes delocalize eigenmodes across the traceless subspace and, through non-unital dissipation, couple them to the identity sector. We show that this delocalization is topological in origin using the spectral localizer~\cite{Loring2015, Loring2017, Cerjan2024}. Crucially, in the BTC phase, eigenmode delocalization toward the identity sector generates initial state independent sourcing of the slowest-decaying oscillatory modes, thereby providing the microscopic mechanism underlying the robust BTC oscillations. More generally, this mechanism persists beyond the BTC regime, but the sourced modes acquire vanishing imaginary parts, resulting in purely relaxational dynamics. 

The paper is organized as follows. In Sec.~\ref{sec:model_and_spherical_tensors}, we introduce the canonical BTC model introduced in Ref.~\cite{Iemini2018} and the spherical tensor basis~\cite{Varshalovich1988, Blum2012, Hecht2000, Nemeth2026}. In Sec.~\ref{sec:nonunital_sourced_dynamics}, we make use of a mapping from Ref.~\cite{Nemeth2026} that allows the BTC model to be interpreted as a NH hopping model, where non-unital dissipation acts as a source for the dynamics of spherical tensor components. In addition, we demonstrate how the propagation of sourced effects is related to eigenmode delocalization. In Sec.~\ref{sec:topological_characterization}, we connect this to non-trivial topology, which we probe numerically via the spectral localizer algorithm~\cite{Cerjan2024, Chadha2024}. Finally, we summarize our results in Sec.~\ref{sec:summary}.

\section{BTC Model and Spherical Tensors}
\label{sec:model_and_spherical_tensors}
\begin{figure}[t]
    \centering
    \includegraphics[width=0.48\textwidth]{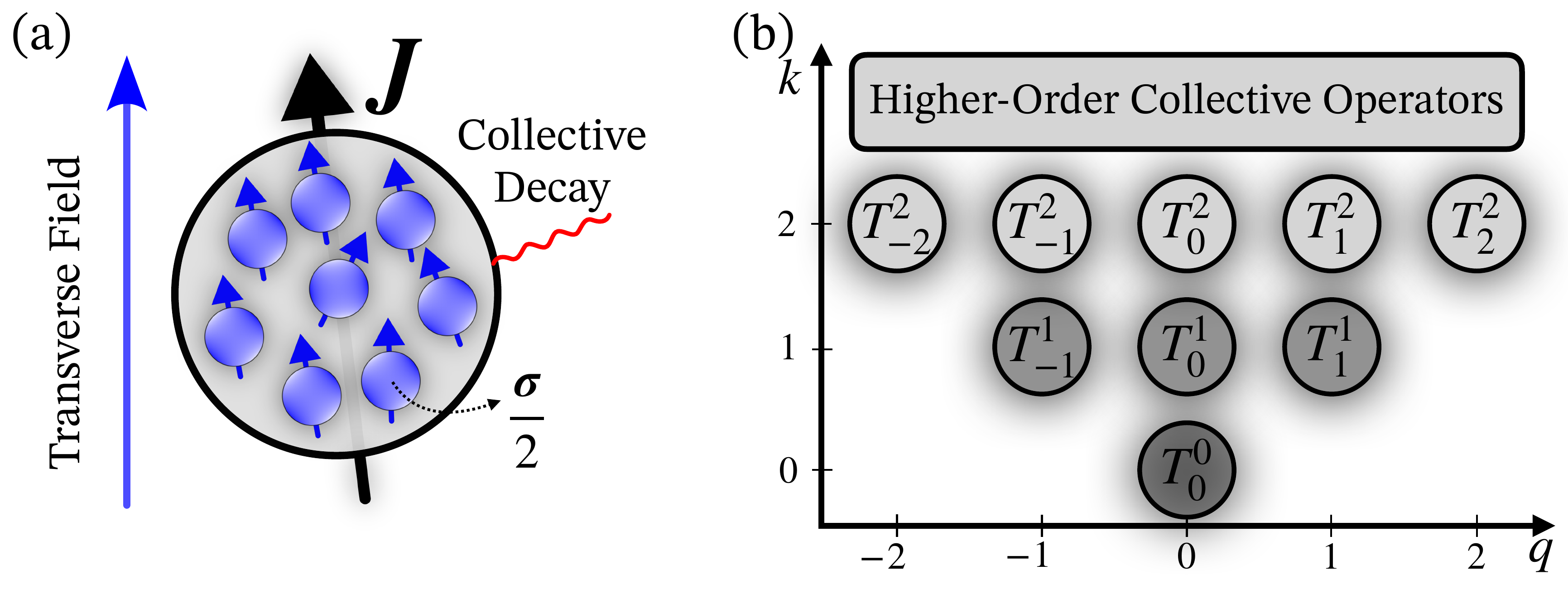}
    \caption{Collective spin system and spherical tensor operators. (a) Schematic diagram of the BTC model in Eq.~(\ref{eq:BTC_master_Eq}), showing the collective spin configuration for the maximally polarized subsector $j=N/2$. (b) Spherical tensor basis [Eq.~(\ref{eq:spherical_tensor_basis})] shown as operator multiplets, where the $k^{\mathrm{th}}$ multiplet contains $(2k+1)$ operators. Here, $T^0_0 \propto \mathbf{1}$ (the identity) and the triplet $(T^1_{-1}, T^1_0, T^1_1)$ is equivalent to $(J_-, J_z, J_+)$ up to normalization factors. The multiplets form a wedge-shaped lattice, whose sites are the spherical tensor operators.} %b) Emergent 1D chain mapping using the spherical tensor operator basis [Eq.~(\ref{eq:non_hermitian_hopping_mapping})]. NH hoppings and on-site decay are labeled via $t_{\pm}(k,q)$ and $\gamma(k,q)$ respectively. c) Internal levels at the $k^{th}$ site with $q=-k, ..., +k$. Coherent transitions are denoted via $w(k,q)$. }
    \label{fig:figure1}
\end{figure}

BTCs can arise in collective spin systems governed by a Markovian Lindblad master equation, $
    \dot{\rho} = \mathcal{L}\rho,$
where $\rho$ is the density operator and $\mathcal{L}$ the Liouvillian superoperator. The canonical BTC model~\cite{Iemini2018} can be written as,
\begin{equation}
    \dot{\rho} = -i[H, \rho] + \dfrac{\Gamma}{2j}\Bigl(J_- \rho J_+ - \dfrac{1}{2}\{J_+ J_-, \rho\}\Bigr),
    \label{eq:BTC_master_Eq}
\end{equation}
where $H=\Omega J_x$ describes uniform coupling to a transverse field, while the dissipator produces collective decay with rate $\Gamma$ [see Fig.~\ref{fig:figure1}\hyperref[fig:figure1]{(a)}]. A microscopic realization is provided by collective operators $J_\alpha$ ($\alpha=x,y,z$) which are built from $N$ underlying spin-$\tfrac{1}{2}$ degrees of freedom. Finally, $J_\pm = J_x \pm iJ_y$ denote the spin ladder operators, and $j$ labels the total spin quantum number. This model can arise in cavity systems with cooperative emission~\cite{Walls1978, Drummond1978, Drummond1980, Schneider2002, Hannukainen2018}. In this setting, the collective spin system acts as a macroscopically large boundary subsystem of a much larger total system, where the surrounding bulk degrees of freedom induce the effective dissipative dynamics on the boundary~\cite{Iemini2018}. The BTC model~[Eq.(\ref{eq:BTC_master_Eq})] possesses a global SU$(2)$ symmetry, implying that the total spin quantum number $j$ is conserved under the dynamics. In the BTC literature, one typically focuses on the maximally symmetric sector, $j=N/2$. However, the formalism developed here applies generally to arbitrary, large spin-$j$ representations, with the corresponding thermodynamic limit given by $j\to\infty$.

The model of Eq.~(\ref{eq:BTC_master_Eq}) exhibits a BTC phase for sufficiently strong coherent driving, $\Gamma/\Omega < 2$, as predicted by mean-field theory~\cite{Iemini2018}. This phase is characterized by persistent oscillations whose decay rates vanish in the thermodynamic limit. Importantly, these oscillations are robust against the choice of initial state, reflecting the intrinsic initial state independence of the BTC phase. A natural basis on which to expand the density matrix is provided by the spherical tensor operators $T^k_q$ \cite{Fano1957, Happer1972, Varshalovich1988,Hecht2000, Blum2012, Nemeth2026},
\begin{equation}
    \rho (t) = \sum_{k=0}^{2j} \sum_{q=-k}^{+k} a_{k,q} (t) T^k_q,
    \label{eq:spherical_tensor_basis}
\end{equation}
where the spherical tensor rank $k$ may take the values $k=0, 1, ..., 2j$ and $q$ is a magnetic quantum number with allowed values $q=-k, ..., +k$. The tensor rank $k$ provides a measure of operator complexity. For example, the $k=0$ sector corresponds to the identity operator, the $k=1$ sector contains the collective spin operators $J_{x,y,z}$, and higher-$k$ sectors describe progressively higher-order collective spin correlations. The magnetic index $q$ labels internal states within a fixed-$k$ sector. For instance, for $k=1$, the spherical tensors ($T^1_0,T^1_1,T^1_{-1}$) are related to the collective spin operators ($J_z,J_+,J_-$), respectively.

\section{NH Hopping Model and Nonunital Sourced Dynamics}
\label{sec:nonunital_sourced_dynamics}
Within this representation, the labels $k$ and $q$ can be interpreted as the coordinates of a two-dimensional wedge shaped lattice as shown in Fig.~\ref{fig:figure1}\hyperref[fig:figure1]{(b)}. Importantly, the action of the superoperator $\mathcal{L}$ can be reinterpreted as hoppings on this lattice.

Expanding $\rho$ in the basis $\{T^k_q\}$ [Eq.~(\ref{eq:spherical_tensor_basis})]
maps Eq.~(\ref{eq:BTC_master_Eq}) to an effective $2$D NH hopping model \cite{Nemeth2026},
\begin{equation}
\begin{aligned}
\dot a_{k,q} =\;&
-\frac{i\Omega}{2}
\sum_{\delta=\pm1}
w_\delta(k,q-\delta)a_{k,q-\delta}
\\
&+
\frac{\Gamma}{2j}
\left[
\sum_{\delta=\pm1}
t_\delta(k-\delta,q, j)a_{k-\delta,q},
-\gamma(k,q)a_{k,q}
\right],
\end{aligned}
\label{eq:non_hermitian_hopping_mapping}
\end{equation}
where $w_\pm(k,q)$, $t_\pm(k,q,j)$ and $\gamma(k,q)$ represent position-dependent hoppings and onsite damping terms. The coherent contribution, $-i[H,\cdot]$, generates nearest-neighbor reciprocal hopping along the $q$-direction via $w_\pm(k,q)$, leading to a symmetric spreading of operator amplitude across the internal $q$-coordinate. In contrast, the dissipative terms generate non-reciprocal hopping along the $k$-direction, with $\vert t_+(k,q, j)\vert \neq \vert t_-(k-1,q,j)\vert$~\cite{Nemeth2026}, which are in general $j$-dependent~\cite{Nemeth2026}. For more detail on the mapping and the resultant hopping and damping terms we refer the Reader to Appendix~\ref{sec:mapping_nh_hopping_model} and Ref.~\cite{Nemeth2026}. Consequently, dissipation induces a directional transport of operator weight through tensor rank space~\cite{Hatano1996,Borgnia2020,Okuma2020}.

This becomes particularly important near the boundary $k=0$, which corresponds to the traceful sector and contains solely the identity operator, while all sectors with $k>0$ are traceless. In this sense, non-unital dynamics allows the identity sector to act as a source for the evolution of the traceless subspace~\cite{Byrd2003, Kimura2003, Bertlmann2008, Byrd2011, Nemeth2026}.

This can be made explicit through the generalized Bloch decomposition~\cite{Byrd2003, Kimura2003, Bertlmann2008, Byrd2011},
\begin{equation}
\rho(t)=\frac{\mathbf1}{d}+\delta\rho(t),
\end{equation}
where $d$ is the Hilbert-space dimension and $\Tr(\delta\rho(t))=0$. The coefficient of $\mathbf1$ is fixed by trace preservation, ensuring $\Tr(\rho(t))=1$ at all times. Within the spherical tensor framework, this decomposition amounts to separating the rank-$0$ identity component from the traceless, higher-rank spherical tensor sectors, which are collectively represented by $\delta \rho(t)$.  In Appendix~\ref{sec:liouvillian_eigenmode_treatment}, we show that the solution for $\delta\rho(t)$ separates into homogeneous and sourced contributions,
\begin{equation}
\begin{split}
\delta \rho(t)
= &
\sum_{\alpha}
e^{\lambda_\alpha t}
\Tr\Bigl(L_\alpha^\dagger \delta\rho(0)\Bigr)
R_\alpha
\\
+&
\sum_{\alpha}
\frac{e^{\lambda_\alpha t}-1}{d}
\Tr\Bigl(L_\alpha^\dagger \mathbf 1\Bigr)
R_\alpha ,
\end{split}
\end{equation}
where $L_\alpha (R_\alpha)$ denote the left (right) Liouvillian eigenoperators with eigenvalue $\lambda_\alpha$. The first term depends explicitly on the initial state through $\delta\rho(0)$, whereas the second term corresponds to an intrinsically generated initial state independent response, sourced directly from the identity operator.

Importantly, in Appendix~\ref{sec:liouvillian_eigenmode_treatment} we show that these sourced responses remain finite in the thermodynamic limit, both for magnetization observables and higher-order intensive spin correlations. This immediately implies two necessary conditions for a finite sourced response: i) the relevant left eigenoperator must possess finite overlap with the identity operator and ii) the corresponding right eigenoperator must simultaneously retain finite overlap with the probe observable. Physically, operator amplitude sourced from the identity sector ($k=0$) must be able to propagate through tensor rank space and reach the observable sector while retaining finite weight. For magnetization observables this requires transport to $k=1$, while higher-order spin correlations require $k$ successive dissipative hoppings [cf. Eq.~\ref{eq:non_hermitian_hopping_mapping}]. Consequently, within the BTC phase, the leading order initial state independent contribution to a degree-$k$ intensive collective spin observable is of order $\Gamma^k$. Numerical evidence of this scaling law is presented in Appendix~\ref{sec:hierarchy_of_spin_correlations}.

These conditions are satisfied when the relevant oscillatory eigenmodes become delocalized along the $k$-direction. As we show below, this delocalization is topologically enforced by a non-trivial local topological index. Consequently, in the remainder of this work we focus on the emergent one-dimensional chain along $k$, with the internal $q$-coordinate suppressed. This chain comprises of \mbox{$(2j+1)$} sites and possesses intrinsic open boundaries at $k=0$ and $k=2j$, as illustrated in Fig.~\ref{fig:figure2}\hyperref[fig:figure2]{(a)}.

\section{Topological Characterization}
\label{sec:topological_characterization}
\subsection{Spectral Localizer Probe}

\begin{figure}[t]
    \centering
    \includegraphics[width=0.48\textwidth]{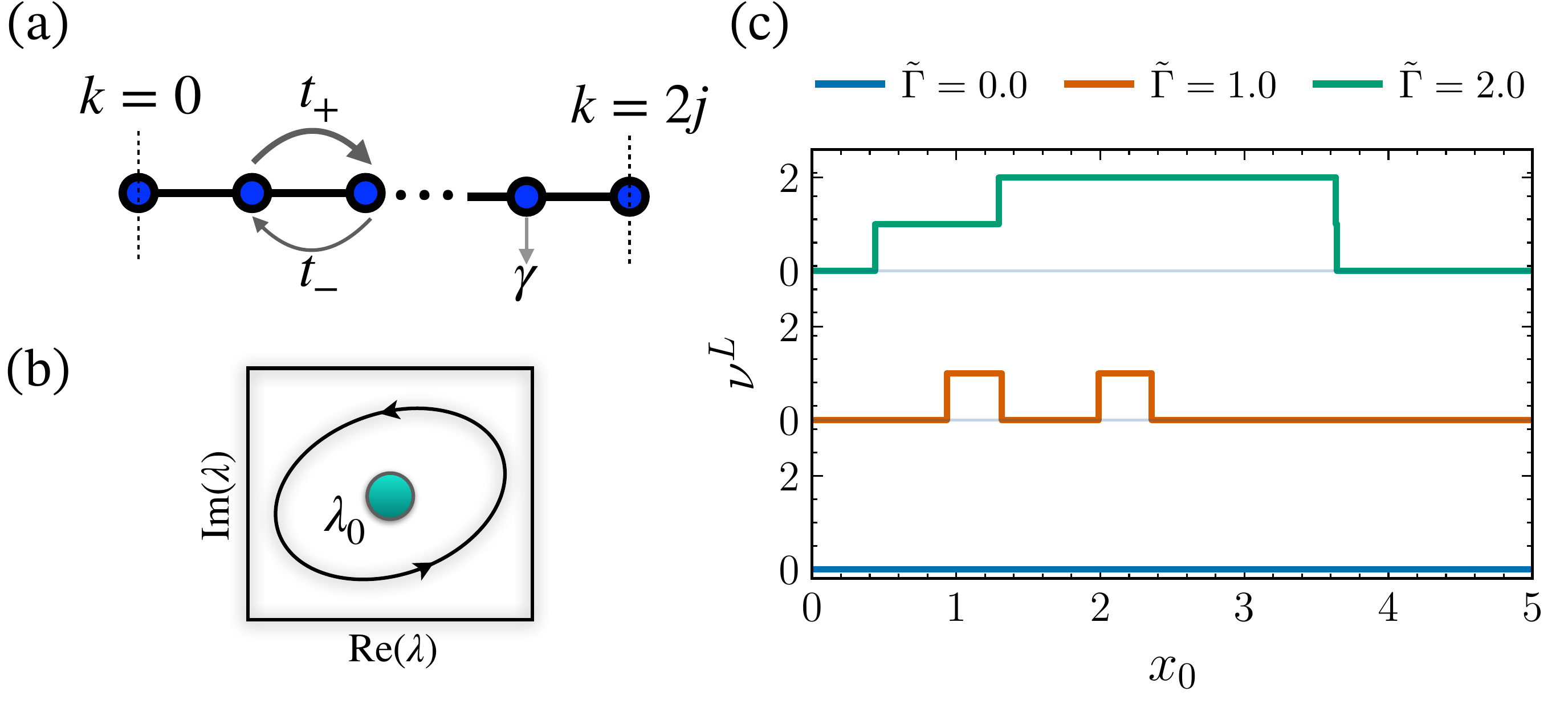}
    \caption{NH hopping model and one-dimensional spectral localizer probe. (a) Effective one-dimensional hopping chain along the probe coordinate $k$, constructed in the spherical tensor basis [Eq.~(\ref{eq:spherical_tensor_basis})]. (b) As $p$ is varied along $S^1$, the map $p \mapsto \mathcal L_U(p)-\lambda_0$ traces a non-contractible loop in the complex plane, signaling a non-trivial winding around the point-gap at $\lambda_0$. (c) Local topological index $\nu^L$ as a function of the sweep coordinate $x_0$, probing the rank chain shown in (a). Results are shown for fixed $\lambda_0$, chosen to coincide with one of the two slowest-decaying oscillatory modes, for $\tilde{\Gamma}\equiv\Gamma/\Omega=0,1.0,2.0$ and $N=10$ spins. Data is truncated to $x_0\leq 5$ for clarity; $\nu^L$ remains zero beyond this range.}
    \label{fig:figure2}
\end{figure}

We classify a local point-gap topology associated with the family of operators \mbox{$x \mapsto \mathcal L(x)-\lambda_0$},
where $x\in[0,2j]$ denotes a continuous coordinate along the emergent one-dimensional chain and $\lambda_0\in\mathbb C$ is a reference complex frequency. In a locally uniform region $U$, where the Liouvillian varies slowly in $x$, the system admits an approximate translation symmetry. One may then introduce a conjugate momentum $p\in S^1$ via a local Fourier transform. As $p$ is varied around the circle, the map $p \mapsto \mathcal L_U(p)-\lambda_0$ defines a closed loop in the space of invertible operators. The winding of this loop provides the local topological invariant. If the map can be continuously deformed to a single point, then it is topologically trivial. If such a deformation is obstructed, the map necessarily exhibits a non-trivial winding, indicating a non-trivial \emph{local point-gap topology} (see Fig.~\ref{fig:figure2}\hyperref[fig:figure2]{(b)}).
The local point-gap topology and its corresponding non-trivial winding can be diagnosed via the spectral localizer framework~\cite{Loring2015,Loring2017, Loring2019,Cerjan2024, Jezequel2025}. This framework converts the initial problem into a Hermitian gap analysis, such that non-trivial windings are encoded into local topological indices.

The spectral localizer provides a position-space diagnostic of topology by asking whether a system can be locally and continuously deformed to an “atomic limit” while preserving both the relevant symmetries and a spectral gap~\cite{Cerjan2024}.
In this atomic limit, the fundamental degrees of freedom are fully decoupled and behave as isolated constituents. Such a deformation implies the existence of a complete, localized Wannier basis and therefore places the system in a topologically trivial class.
Conversely, when no such local deformation exists, the obstruction signals non-trivial topology—in the present context, a non-trivial point-gap winding of the spectrum. The spectral localizer formulates this obstruction directly in a position-resolved manner, enabling a local probe of topology even in systems without translational invariance. 

In the following, we analyze the resulting one-dimensional chain shown in Fig.~\ref{fig:figure2}\hyperref[fig:figure2]{(a)}. Accordingly, we introduce a position superoperator $\mathcal{X}$ as the projector onto $k$~\cite{Nemeth2026}, whose eigenvalues correspond to the site labels $k=0,1,\ldots,2j$. To probe the local topology, we define the spectral localizer at coordinate $x_0$ and complex frequency $\lambda_0$ as
\begin{equation}
\begin{split}
L_{(x_0,\lambda_0)}(\mathcal{L},\mathcal{X}) =&
\mathrm{Re}(\mathcal{L}-\lambda_0\mathbb{I}) \otimes \sigma_x
+ \mathrm{Im}(\mathcal{L}-\lambda_0\mathbb{I}) \otimes \sigma_y \\
&+ \kappa(\mathcal{X}-x_0\mathbb{I}) \otimes \sigma_z ,
\label{eq:spectral_localizer}
\end{split}
\end{equation}
where $\kappa$ controls the relative strength of spatial and spectral terms, and the Pauli matrices $\sigma_{x,y,z}$ enforce an orthogonal embedding of spectral and positional directions \cite{Cerjan2024}.

This construction converts the non-trivial point-gap winding into an integer-valued local topological invariant,
\begin{equation}
\nu^{L}_{(x_0,\lambda_0)}
= \frac{1}{2}\mathrm{sig}\Bigl[L_{(x_0,\lambda_0)}\Bigr],
\end{equation}
where the signature is defined as the difference between the number of positive and negative eigenvalues. The corresponding localizer gap is
\begin{equation}
    \mu_{(x_0,\lambda_0)}=\min\Bigl\{\,|\lambda|:\lambda\in \mathrm{spec}\bigl(L_{(x_0,\lambda_0)}\bigr)\Bigr\},
\end{equation}
where $\mathrm{spec}(L_{(x_0,\lambda_0)})$ denotes the spectrum of the (Hermitian) localizer evaluated at coordinate $x_0$ and complex reference frequency $\lambda_0$. The index $\nu^L$ is stable under any continuous deformation that preserves $\mu_{(x_0,\lambda_0)}>0$; it can change only when the localizer gap closes, i.e., when an eigenvalue of $L_{(x_0,\lambda_0)}$ crosses zero, signalling a topological transition \cite{Cerjan2024}.

A non-zero value of $\nu^{L}_{(x_0,\lambda_0)}$ implies that $\mathcal{L}$ and $\mathcal{X}$ cannot be smoothly deformed to a mutually commuting, locally trivial limit without closing the gap $\mu_{(x_0, \lambda_0)}$ \cite{Cerjan2024}. Consequently, there exists a topological obstruction to constructing operator modes that are simultaneously localized near position $x_0$ and sharply defined at complex frequency $\lambda_0$. Physically, this constitutes a no-go statement: the dynamics near $(x_0,\lambda_0)$ cannot be reduced to independent, localized operator modes. In terms of the spherical tensor representation, this distinction has a clear interpretation. When $\nu^{L}_{(x_0,\lambda_0)}=0$, Liouvillian eigenmodes can be constructed whose expansion coefficients $a_{k,q}$ are sharply peaked around a single tensor rank $k_0$, corresponding to localized operator modes in operator space. By contrast, when $\nu^{L}_{(x_0,\lambda_0)}\neq0$, such localization is forbidden: any eigenmode with complex frequency near $\lambda_0$ must necessarily acquire support over an extended range of tensor ranks. The localizer index therefore diagnoses a fundamental incompatibility between localization in operator space and localization in complex frequency, enforcing delocalization of Liouvillian eigenmodes across spherical tensor sectors.

\begin{figure*}[ht]
    \centering
    \includegraphics[width=0.99\linewidth]{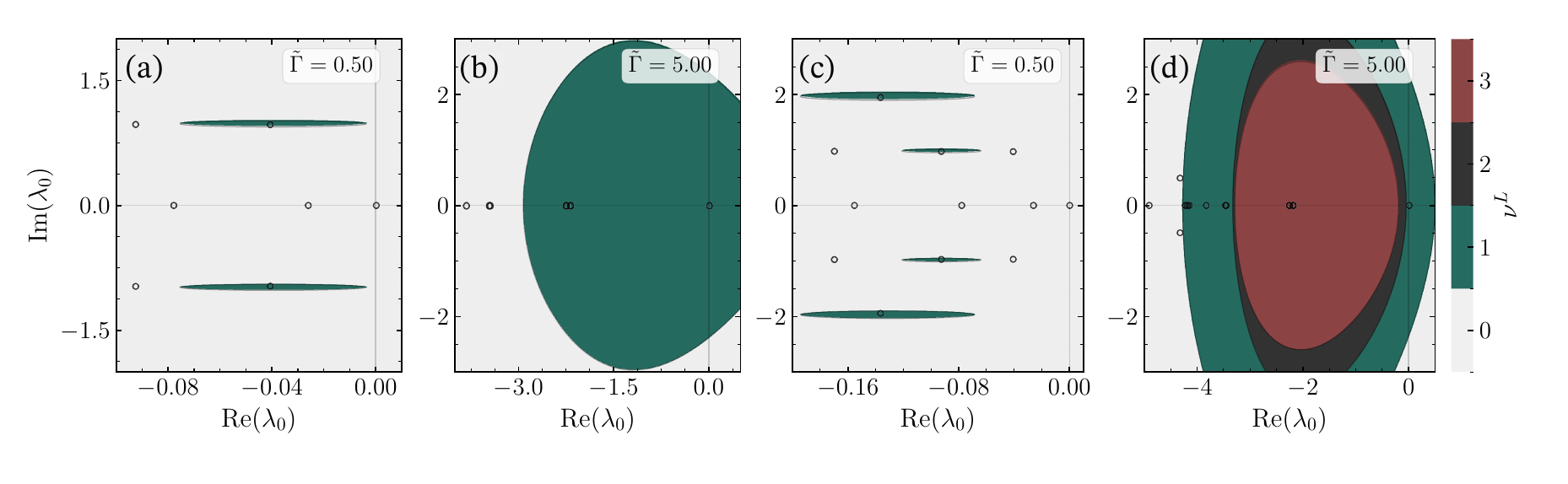}
    \caption{Topological islands in the complex-frequency plane. For fixed operator space coordinate $x_0=1$ ((a) and (b)) and $x_0=2$ ((c) and (d)), we show the complex-frequency-resolved local index $\nu^L(\lambda_0)$ for two representative cases. (a) and (c) correspond to $\tilde{\Gamma}\equiv \Gamma/\Omega = 0.50$ in the BTC regime for $N=10$ spins. (b) and (d) show the same system at $\tilde{\Gamma}=5.0$, corresponding to the deep non-BTC regime. Hollow circular markers denote the eigenvalues of the Liouvillian $\mathcal{L}$. Colour shading indicates the local value of $\nu^L$, revealing spatially resolved regions of non-trivial topology in the complex-frequency plane.}
    \label{fig:topological_islands_1}
\end{figure*}

\subsection{Spectral Signatures}
\label{sec:spectral_signatures}
We begin by fixing $\lambda_0$ to one member of the slowest-decaying oscillatory eigenmode pair governing the dominant BTC oscillations. Scanning the localizer as a function of $x_0$, in Fig.~\ref{fig:figure2}\hyperref[fig:figure2]{(c)} we observe the emergence and subsequent merging of position-dependent topological domains as the dissipation strength $\Gamma$ increases. These domains are local regions with non-zero topological index. Importantly, the ratio $t_+(k,q,j)/t_-(k,q,j)$ remains independent of $\Gamma$ in the BTC model, implying that non-reciprocal transport is present for any non-zero dissipation. Consequently, both the apparent onset and merging of topological domains should be interpreted as the NH transport along the $k$-direction becoming progressively resolved by the finite-resolution localizer probe, rather than as a microscopic onset of non-reciprocal transport itself. Note that this interpretation is specific to BTC models of the form in Eq.~(\ref{eq:BTC_master_Eq}). In contrast, non-BTC models, such as purely collective precession systems, instead generate a $q$-dependent on-site damping rather than directional transport along $k$~\cite{Nemeth2026}.

Rather than fixing a complex frequency and probing the operator space chain, we may adopt the complementary perspective of fixing an operator space coordinate and resolving the entire complex frequency plane, see Fig.~\ref{fig:topological_islands_1}. This allows us to determine which Liouvillian eigenmodes experience a topological obstruction to localization in tensor rank. Performing this spectral sweep reveals the emergence of isolated regions in the complex $\lambda$-plane where the localizer index is non-zero, forming topological “islands” in spectral space.

%These islands act as \emph{local Chern-like markers} in spectral space: they signal a topological obstruction associated with modes at those complex frequencies when projected onto the chosen operator sector.

Fixing $x_0=1$ to probe the magnetization sector, we find that these topological islands host the slowest-decaying oscillatory modes of the Liouvillian spectrum, Fig.~\ref{fig:topological_islands_1}\hyperref[fig:topological_islands_1]{a)}. These modes dominate the long-time dynamics and are responsible for the persistent oscillations characteristic of the BTC phase \cite{Iemini2018}. Since they necessarily carry a non-zero $\nu^L$, they constitute \emph{topological oscillatory modes}: their existence and spectral isolation are protected against local, continuous deformations of $\mathcal{L}$. This provides a natural explanation for the robustness of BTCs, directly linking their long-lived oscillatory behavior to an underlying, observable-resolved operator space topology.

While Fig.~\ref{fig:topological_islands_1}\hyperref[fig:topological_islands_1]{(a)} corresponds to the BTC phase, Fig.~\ref{fig:topological_islands_1}\hyperref[fig:topological_islands_1]{(b)} illustrates the non-BTC regime. This reveals an important feature: for any non-zero $\Gamma$, directional transport along the $k$-direction, and hence non-trivial local topology, persists even beyond the BTC phase. Consequently, the BTC phase transition itself does not correspond to a topological transition. Instead, the phase transition arises from the competition between coherent and dissipative dynamics, leading to the emergence of exceptional points in finite-size systems~\cite{Iemini2018,Nakanishi2023,Nemethb2025}.

Importantly, in the non-BTC regime, the eigenmodes associated with the topologically non-trivial spectral islands acquire vanishing imaginary parts, such that the corresponding topological modes become purely relaxational. As a consequence, the robust initial state independent response persists beyond the BTC phase, but is governed by decaying rather than oscillatory dynamics.

We further demonstrate the existence of topological modes beyond magnetization observables in Fig.~\ref{fig:topological_islands_1}\hyperref[fig:topological_islands_1]{(c)-(d)}. In particular, probing the quadratic collective spin sector by fixing $x_0=2$, we identify two distinct pairs of oscillatory eigenmodes with different frequencies. Consequently, the resulting robust dynamics arises from a superposition of these modes, producing a richer interference structure. More generally, as anticipated above, these robust initial state independent features persist across higher-order collective correlation functions, albeit with progressively suppressed amplitudes, see Appendix~\ref{sec:hierarchy_of_spin_correlations} for further details.

\subsection{Eigenmode Delocalization}
\label{sec:eigenmode_delocalization}

\begin{figure*}[ht!]
    \centering
    \includegraphics[width=\linewidth]{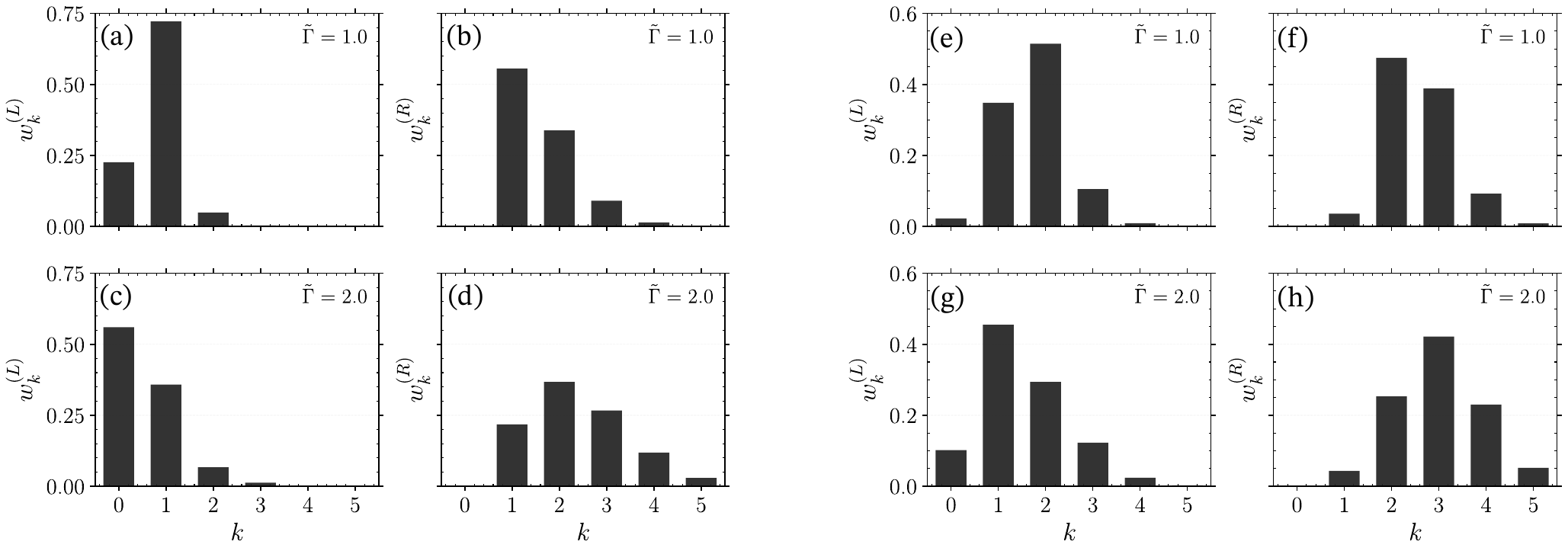}
    \caption{Delocalization of Lindbladian eigenmodes. (a)-(d) correspond to a slowest-decaying oscillatory eigenmode and (e)-(h) correspond to a subleading slow-decaying oscillatory eigenmode. Rank-resolved mode weights $w_k^{(L/R)}$ are shown for the corresponding left and right eigenvectors. The first and second rows correspond to $\tilde{\Gamma}\equiv \Gamma/\Omega = 1.0$ and $2.0$, respectively, with $N=5$ spins throughout.}
    \label{fig:eigenmode_delocalization1}
\end{figure*}

We provide further numerical evidence for the eigenmode delocalization discussed above. To quantify these delocalization effects, we plot the rank-resolved weight profiles $w_k^{(L/R)}$ of the dominant BTC mode in Fig.~\ref{fig:eigenmode_delocalization1}. A right (left) eigenvector is expanded in the vectorized spherical tensor basis as
\begin{equation}
\ket{r^{(\alpha)}}\rangle = \sum_{k,q} c^{(\alpha)}_{kq}\ket{k,q}\rangle .
\end{equation}
The basis states $\ket{k\,q}\rangle$ are orthonormal under the Hilbert--Schmidt inner product~\cite{Nemeth2026}. The rank-resolved weights are then defined as
\begin{equation}
w_k^{(L/R)} = \sum_q \left|c^{(L/R)}_{kq}\right|^2 ,
\end{equation}
which quantify the occupation of each rank-$k$ sector by the corresponding eigenvector. 

Fig.~\ref{fig:eigenmode_delocalization1}\hyperref[fig:eigenmode_delocalization1]{(a)-(d)} shows that for the slowest decaying oscillatory eigenmode the corresponding left eigenvector possesses a finite overlap with the identity sector ($k=0$), while the corresponding right eigenvector exhibits a finite overlap with the magnetization sector ($k=1$). Consequently, this eigenmode contributes to the inhomogeneous component of the magnetization dynamics. We highlight analogous features for a subleading slow-decaying oscillatory eigenmode, which contributes to the dynamics of quadratic collective spin operators in Fig.~\ref{fig:eigenmode_delocalization1}\hyperref[fig:eigenmode_delocalization1]{(e)-(h)}. In this case, however, the overlap with the identity sector is substantially reduced due to the additional factor of $\Gamma$ acquired through an extra dissipative hopping process required to propagate from $k=0$ to $k=2$. Furthermore, at weaker dissipation, this mode contributes predominantly to the $k=2$ probe sector.

This behavior highlights that the relevant topology is inherently complex-frequency resolved. Rather than defining a single global topological phase, the localizer probes the obstruction structure associated with a fixed spectral point $\lambda_0$. As a result, different eigenmodes experience distinct patterns of topological domains along the operator space chain. Operator space delocalization is therefore mode-dependent, with each eigenmode exhibiting a characteristic spatial profile across tensor rank sectors.

\section{Summary and Outlook}
\label{sec:summary}
By expressing the collective spin Lindbladian in a spherical tensor basis, we mapped the canonical BTC model [Eq.~(\ref{eq:BTC_master_Eq})] onto an effective NH hopping problem in operator space [Eq.~(\ref{eq:non_hermitian_hopping_mapping})]. Within this framework, non-unital dissipation acts as a boundary source at $k=0$, driving dynamics into the traceless operator subspace, while NH topology constrains the propagation of operator amplitude toward physical observables. In particular, a non-trivial local winding number enforces the delocalization of Liouvillian eigenmodes across tensor rank space, ensuring finite overlap with both source and probe sectors. As a consequence, the system generically exhibits robust initial state independent sourced dynamics for any non-zero dissipation strength: oscillatory in the BTC regime and relaxational outside it.

These effects extend beyond magnetization observables and persist throughout the hierarchy of collective spin correlations, with sourced response amplitudes scaling at least as $\Gamma^k$ for degree-$k$ intensive operators. Moreover, maximally mixed initial states isolate the purely sourced contribution directly, providing a clear signature of the beyond mean-field non-unital dynamics. Our results therefore provide a microscopic mechanism for robust initial state independent behavior and yield experimentally testable predictions for intrinsically collective platforms such as molecular and nuclear spin gases~\cite{Huang2025} and atom-cavity systems~\cite{Kongkhambut2022}.

More broadly, our work suggests a general connection between non-unital dissipation, NH topology and robust dynamical responses in open systems. Since non-unital quantum channels generically lead to affine dynamics~\cite{Byrd2003, Kimura2003, Bertlmann2008, Byrd2011}, related phenomena may also emerge in Floquet systems~\cite{Koch2025}, quantum feedback control~\cite{Nakagawa2025}, and classical stochastic dynamics~\cite{Nakagawa2026}.

Moreover, our results motivate the exploration of operator space models that support genuinely higher-dimensional topological structures. While the BTCs studied here can be understood as effective one-dimensional chains labeled by tensor rank $k$, with motion in the magnetic quantum number $q$ occurring locally at each site, our findings suggest the possibility of realizing local Chern markers in the full $(k,q)$ operator space geometry. Such behavior is expected to arise in models where these coordinates cannot be treated independently, opening new avenues for topological classification and control of many-body open quantum dynamics.

\begin{acknowledgements}
We acknowledge the use of the QuTiP library~\cite{Lambert2024}. D.N. acknowledges support from The University of Manchester under the Dean's Doctoral Scholarship. A.P. acknowledges support from the Leverhulme Trust under the grant agreement RPG-2023-253. R.-J.S. acknowledges funding from an EPSRC ERC underwrite Grant No.~EP/X025829/1, and a Royal Society exchange Grant No. IES/R1/221060, as well as Trinity College, Cambridge. Code and data in support of the findings of this paper are available on GitHub~\cite{nemeth2026code}. D.N. acknowledges the use of ChatGPT (OpenAI, GPT-5) for cross-verification of mathematical derivations in the Appendices, assistance with editing and refactoring code and in the numerical implementation of an LDL accelerated spectral localizer algorithm, which are discussed in Refs.~\cite{Cerjan2024, Chadha2024}. All suggestions were reviewed, tested against a direct implementation and validated by D.N. who takes full responsibility for its content.
\end{acknowledgements}

\bibliography{bibliography}

@article{Wilczek2012,
  title = {{Quantum} {Time} {Crystals}},
  author = {Wilczek, Frank},
  journal = {Phys. Rev. Lett.},
  volume = {109},
  issue = {16},
  pages = {160401},
  numpages = {5},
  year = {2012},
  month = {Oct},
  publisher = {American Physical Society},
  doi = {10.1103/PhysRevLett.109.160401},
  url = {https://link.aps.org/doi/10.1103/PhysRevLett.109.160401}
}

@article{Bruno2013,
  title = {{Impossibility} of {Spontaneously Rotating Time Crystals: A No-Go Theorem}},
  author = {Bruno, Patrick},
  journal = {Phys. Rev. Lett.},
  volume = {111},
  issue = {7},
  pages = {070402},
  numpages = {5},
  year = {2013},
  month = {Aug},
  publisher = {American Physical Society},
  doi = {10.1103/PhysRevLett.111.070402},
  url = {https://link.aps.org/doi/10.1103/PhysRevLett.111.070402}
}

@article{Watanabe2015,
  title = {{Absence} of {Quantum Time Crystals}},
  author = {Watanabe, Haruki and Oshikawa, Masaki},
  journal = {Phys. Rev. Lett.},
  volume = {114},
  issue = {25},
  pages = {251603},
  numpages = {5},
  year = {2015},
  month = {Jun},
  publisher = {American Physical Society},
  doi = {10.1103/PhysRevLett.114.251603},
  url = {https://link.aps.org/doi/10.1103/PhysRevLett.114.251603}
}

@article{Sacha2015,
  title = {Modeling spontaneous breaking of time-translation symmetry},
  author = {Sacha, Krzysztof},
  journal = {Phys. Rev. A},
  volume = {91},
  issue = {3},
  pages = {033617},
  numpages = {5},
  year = {2015},
  month = {Mar},
  publisher = {American Physical Society},
  doi = {10.1103/PhysRevA.91.033617},
  url = {https://link.aps.org/doi/10.1103/PhysRevA.91.033617}
}

@article{Khemani2016,
  title = {{Phase Structure} of {Driven Quantum Systems}},
  author = {Khemani, Vedika and Lazarides, Achilleas and Moessner, Roderich and Sondhi, S. L.},
  journal = {Phys. Rev. Lett.},
  volume = {116},
  issue = {25},
  pages = {250401},
  numpages = {6},
  year = {2016},
  month = {Jun},
  publisher = {American Physical Society},
  doi = {10.1103/PhysRevLett.116.250401},
  url = {https://link.aps.org/doi/10.1103/PhysRevLett.116.250401}
}

@misc{Khemani2019,
      title={A {Brief} {History} of {Time} {Crystals}}, 
      author={Vedika Khemani and Roderich Moessner and S. L. Sondhi},
      year={2019},
      eprint={1910.10745},
      archivePrefix={arXiv},
      primaryClass={cond-mat.str-el},
      url={https://arxiv.org/abs/1910.10745}, 
}

@article{Else2016,
  title = {{Floquet Time Crystals}},
  author = {Else, Dominic V. and Bauer, Bela and Nayak, Chetan},
  journal = {Phys. Rev. Lett.},
  volume = {117},
  issue = {9},
  pages = {090402},
  numpages = {5},
  year = {2016},
  month = {Aug},
  publisher = {American Physical Society},
  doi = {10.1103/PhysRevLett.117.090402},
  url = {https://link.aps.org/doi/10.1103/PhysRevLett.117.090402}
}

@article{Keyserlingk2016,
  title = {Absolute stability and spatiotemporal long-range order in {Floquet} systems},
  author = {von Keyserlingk, C. W. and Khemani, Vedika and Sondhi, S. L.},
  journal = {Phys. Rev. B},
  volume = {94},
  issue = {8},
  pages = {085112},
  numpages = {11},
  year = {2016},
  month = {Aug},
  publisher = {American Physical Society},
  doi = {10.1103/PhysRevB.94.085112},
  url = {https://link.aps.org/doi/10.1103/PhysRevB.94.085112}
}

@article{Else2017,
  title = {{Prethermal Phases} of {Matter Protected} by {Time-Translation Symmetry}},
  author = {Else, Dominic V. and Bauer, Bela and Nayak, Chetan},
  journal = {Phys. Rev. X},
  volume = {7},
  issue = {1},
  pages = {011026},
  numpages = {21},
  year = {2017},
  month = {Mar},
  publisher = {American Physical Society},
  doi = {10.1103/PhysRevX.7.011026},
  url = {https://link.aps.org/doi/10.1103/PhysRevX.7.011026}
}

@article{Yao2017,
  title = {{Discrete Time Crystals: Rigidity, Criticality}, and {Realizations}},
  author = {Yao, N. Y. and Potter, A. C. and Potirniche, I.-D. and Vishwanath, A.},
  journal = {Phys. Rev. Lett.},
  volume = {118},
  issue = {3},
  pages = {030401},
  numpages = {6},
  year = {2017},
  month = {Jan},
  publisher = {American Physical Society},
  doi = {10.1103/PhysRevLett.118.030401},
  url = {https://link.aps.org/doi/10.1103/PhysRevLett.118.030401}
}

@article{Lazarides2017,
  title = {Fate of a discrete time crystal in an open system},
  author = {Lazarides, Achilleas and Moessner, Roderich},
  journal = {Phys. Rev. B},
  volume = {95},
  issue = {19},
  pages = {195135},
  numpages = {10},
  year = {2017},
  month = {May},
  publisher = {American Physical Society},
  doi = {10.1103/PhysRevB.95.195135},
  url = {https://link.aps.org/doi/10.1103/PhysRevB.95.195135}
}

@article{Sacha2018,
doi = {10.1088/1361-6633/aa8b38},
url = {https://doi.org/10.1088/1361-6633/aa8b38},
year = {2017},
month = {nov},
publisher = {IOP Publishing},
volume = {81},
number = {1},
pages = {016401},
author = {Sacha, Krzysztof and Zakrzewski, Jakub},
title = {Time crystals: a review},
journal = {Reports on Progress in Physics}
}

@article{Zaletel2023,
  title = {Colloquium: Quantum and classical discrete time crystals},
  author = {Zaletel, Michael P. and Lukin, Mikhail and Monroe, Christopher and Nayak, Chetan and Wilczek, Frank and Yao, Norman Y.},
  journal = {Rev. Mod. Phys.},
  volume = {95},
  issue = {3},
  pages = {031001},
  numpages = {34},
  year = {2023},
  month = {Jul},
  publisher = {American Physical Society},
  doi = {10.1103/RevModPhys.95.031001},
  url = {https://link.aps.org/doi/10.1103/RevModPhys.95.031001}
}

@article{Else2020,
	author = {Else, Dominic V. and Monroe, Christopher and Nayak, Chetan and Yao, Norman Y.},
	doi = {https://doi.org/10.1146/annurev-conmatphys-031119-050658},
	issn = {1947-5462},
	journal = {Annual Review of Condensed Matter Physics},
	number = {Volume 11, 2020},
	pages = {467-499},
	publisher = {Annual Reviews},
	title = {{Discrete Time Crystals}},
	type = {Journal Article},
	url = {https://www.annualreviews.org/content/journals/10.1146/annurev-conmatphys-031119-050658},
	volume = {11},
	year = {2020}}

@article{Zhang2017,
	author = {Zhang, J. and Hess, P. W. and Kyprianidis, A. and Becker, P. and Lee, A. and Smith, J. and Pagano, G. and Potirniche, I. -D. and Potter, A. C. and Vishwanath, A. and Yao, N. Y. and Monroe, C.},
	date = {2017/03/01},
	doi = {10.1038/nature21413},
	id = {Zhang2017},
	isbn = {1476-4687},
	journal = {Nature},
	number = {7644},
	pages = {217--220},
	title = {Observation of a discrete time crystal},
	url = {https://doi.org/10.1038/nature21413},
	volume = {543},
	year = {2017}}

@article{Iemini2018,
  title = {{Boundary Time Crystals}},
  author = {Iemini, F. and Russomanno, A. and Keeling, J. and Schir\`o, M. and Dalmonte, M. and Fazio, R.},
  journal = {Phys. Rev. Lett.},
  volume = {121},
  issue = {3},
  pages = {035301},
  numpages = {6},
  year = {2018},
  month = {Jul},
  publisher = {American Physical Society},
  doi = {10.1103/PhysRevLett.121.035301},
  url = {https://link.aps.org/doi/10.1103/PhysRevLett.121.035301}
}

@article{Seibold2020,
  title = {Dissipative time crystal in an asymmetric nonlinear photonic dimer},
  author = {Seibold, Kilian and Rota, Riccardo and Savona, Vincenzo},
  journal = {Phys. Rev. A},
  volume = {101},
  issue = {3},
  pages = {033839},
  numpages = {9},
  year = {2020},
  month = {Mar},
  publisher = {American Physical Society},
  doi = {10.1103/PhysRevA.101.033839},
  url = {https://link.aps.org/doi/10.1103/PhysRevA.101.033839}
}

@article{Lledo2020,
doi = {10.1088/1367-2630/ab9ae3},
url = {https://dx.doi.org/10.1088/1367-2630/ab9ae3},
year = {2020},
month = {jul},
publisher = {IOP Publishing},
volume = {22},
number = {7},
pages = {075002},
author = {Lledó, Cristóbal and Szymańska, Marzena H},
title = {A dissipative time crystal with or without {Z2} symmetry breaking},
journal = {New Journal of Physics}
}

@article{Passarelli2022,
  title = {{Dissipative} time crystals with long-range {Lindbladians}},
  author = {Passarelli, Gianluca and Lucignano, Procolo and Fazio, Rosario and Russomanno, Angelo},
  journal = {Phys. Rev. B},
  volume = {106},
  issue = {22},
  pages = {224308},
  numpages = {13},
  year = {2022},
  month = {Dec},
  publisher = {American Physical Society},
  doi = {10.1103/PhysRevB.106.224308},
  url = {https://link.aps.org/doi/10.1103/PhysRevB.106.224308}
}

@article{Carollo2021,
  title = {Exactness of {Mean-Field Equations} for {Open Dicke Models} with an {Application} to {Pattern Retrieval Dynamics}},
  author = {Carollo, Federico and Lesanovsky, Igor},
  journal = {Phys. Rev. Lett.},
  volume = {126},
  issue = {23},
  pages = {230601},
  numpages = {6},
  year = {2021},
  month = {Jun},
  publisher = {American Physical Society},
  doi = {10.1103/PhysRevLett.126.230601},
  url = {https://link.aps.org/doi/10.1103/PhysRevLett.126.230601}
}

@article{Carollo2022,
  title = {Exact solution of a boundary time-crystal phase transition: Time-translation symmetry breaking and non-Markovian dynamics of correlations},
  author = {Carollo, Federico and Lesanovsky, Igor},
  journal = {Phys. Rev. A},
  volume = {105},
  issue = {4},
  pages = {L040202},
  numpages = {6},
  year = {2022},
  month = {Apr},
  publisher = {American Physical Society},
  doi = {10.1103/PhysRevA.105.L040202},
  url = {https://link.aps.org/doi/10.1103/PhysRevA.105.L040202}
}

@article{Booker2020,
doi = {10.1088/1367-2630/ababc4},
url = {https://doi.org/10.1088/1367-2630/ababc4},
year = {2020},
month = {aug},
publisher = {IOP Publishing},
volume = {22},
number = {8},
pages = {085007},
author = {Booker, Cameron and Buča, Berislav and Jaksch, Dieter},
title = {Non-stationarity and dissipative time crystals: spectral properties and finite-size effects},
journal = {New Journal of Physics}}

@article{Nakanishi2023,
  title = {Dissipative time crystals originating from parity-time symmetry},
  author = {Nakanishi, Yuma and Sasamoto, Tomohiro},
  journal = {Phys. Rev. A},
  volume = {107},
  issue = {1},
  pages = {L010201},
  numpages = {7},
  year = {2023},
  month = {Jan},
  publisher = {American Physical Society},
  doi = {10.1103/PhysRevA.107.L010201},
  url = {https://link.aps.org/doi/10.1103/PhysRevA.107.L010201}
}

@misc{Nemethb2025,
      title={Solving boundary time crystals via the superspin method}, 
      author={Dominik Nemeth and Alessandro Principi and Ahsan Nazir},
      year={2025},
      eprint={2507.06998},
      archivePrefix={arXiv},
      primaryClass={quant-ph},
      url={https://arxiv.org/abs/2507.06998}, 
}

@article{Piccito2021,
  title = {Symmetries and conserved quantities of boundary time crystals in generalized spin models},
  author = {Piccitto, Giulia and Wauters, Matteo and Nori, Franco and Shammah, Nathan},
  journal = {Phys. Rev. B},
  volume = {104},
  issue = {1},
  pages = {014307},
  numpages = {17},
  year = {2021},
  month = {Jul},
  publisher = {American Physical Society},
  doi = {10.1103/PhysRevB.104.014307},
  url = {https://link.aps.org/doi/10.1103/PhysRevB.104.014307}
}

@article{Fiorelli2023,
doi = {10.1088/1367-2630/ace470},
url = {https://doi.org/10.1088/1367-2630/ace470},
year = {2023},
month = {aug},
publisher = {IOP Publishing},
volume = {25},
number = {8},
pages = {083010},
author = {Fiorelli, Eliana and Müller, Markus and Lesanovsky, Igor and Carollo, Federico},
title = {Mean-field dynamics of open quantum systems with collective operator-valued rates: validity and application},
journal = {New Journal of Physics},
}

@article{Mukherjee2024,
  title = {Symmetries and correlations in continuous time crystals},
  author = {Mukherjee, Ankan and Ibrahim, Yeshma and Hajdu\ifmmode \check{s}\else \v{s}\fi{}ek, Michal and Vinjanampathy, Sai},
  journal = {Phys. Rev. A},
  volume = {110},
  issue = {1},
  pages = {012220},
  numpages = {10},
  year = {2024},
  month = {Jul},
  publisher = {American Physical Society},
  doi = {10.1103/PhysRevA.110.012220},
  url = {https://link.aps.org/doi/10.1103/PhysRevA.110.012220}
}

@article{Lourenco2022,
  title = {{Genuine multipartite correlations in a boundary time crystal}},
  author = {Louren\ifmmode \mbox{\c{c}}\else \c{c}\fi{}o, Ant\^onio C. and Prazeres, Luis Fernando dos and Maciel, Thiago O. and Iemini, Fernando and Duzzioni, Eduardo I.},
  journal = {Phys. Rev. B},
  volume = {105},
  issue = {13},
  pages = {134422},
  numpages = {13},
  year = {2022},
  month = {Apr},
  publisher = {American Physical Society},
  doi = {10.1103/PhysRevB.105.134422},
  url = {https://link.aps.org/doi/10.1103/PhysRevB.105.134422}
}

@misc{Liu2025,
      title={{Boundary Time Crystals: Beyond Mean-Field Theory}}, 
      author={Zeping Liu and Yaotian Li and Zhaoyu Fei and Xiaoguang Wang},
      year={2025},
      eprint={2510.03028},
      archivePrefix={arXiv},
      primaryClass={quant-ph},
      url={https://arxiv.org/abs/2510.03028}, 
}

@article{Huang2025,
	author = {Huang, Ying and Wang, Tishuo and Yin, Haochuan and Jiang, Min and Luo, Zhihuang and Peng, Xinhua},
	date = {2025/10/23},
	doi = {10.1038/s41467-025-64413-y},
	id = {Huang2025},
	isbn = {2041-1723},
	journal = {Nature Communications},
	number = {1},
	pages = {9375},
	title = {Observation of continuous time crystals and quasi-crystals in spin gases},
	url = {https://doi.org/10.1038/s41467-025-64413-y},
	volume = {16},
	year = {2025}}

@article{Kongkhambut2022,
   title={Observation of a continuous time crystal},
   volume={377},
   ISSN={1095-9203},
   url={http://dx.doi.org/10.1126/science.abo3382},
   DOI={10.1126/science.abo3382},
   number={6606},
   journal={Science},
   publisher={American Association for the Advancement of Science (AAAS)},
   author={Kongkhambut, Phatthamon and Skulte, Jim and Mathey, Ludwig and Cosme, Jayson G. and Hemmerich, Andreas and Keßler, Hans},
   year={2022},
   month=Aug, pages={670–673} }

@article{Kruthoff2017,
	Author = {Kruthoff, Jorrit and de Boer, Jan and van Wezel, Jasper and Kane, Charles L. and Slager, {Robert-Jan}},
	Doi = {10.1103/PhysRevX.7.041069},
	Issue = {4},
	Journal = {Phys. Rev. X},
	Month = {Dec},
	Numpages = {23},
	Pages = {041069},
	Publisher = {American Physical Society},
	title = {{Topological Classification of Crystalline Insulators through Band Structure Combinatorics}},
	Url = {https://link.aps.org/doi/10.1103/PhysRevX.7.041069},
	Volume = {7},
	Year = {2017}}

@article{Bradlyn2017,
	Author = {Bradlyn, Barry and Elcoro, L. and Cano, Jennifer and Vergniory, M. G. and Wang, Zhijun and Felser, C. and Aroyo, M. I. and Bernevig, B. Andrei},
	Date = {2017/07/19/online},
	Day = {19},
	Journal = {Nature},
	L3 = {10.1038/nature23268; https://www.nature.com/articles/nature23268#supplementary-information},
	M3 = {Article},
	Month = {07},
	Pages = {298},
	Publisher = {Macmillan Publishers Limited, part of Springer Nature. All rights reserved. SN -},
	title = {{Topological quantum chemistry}},
	Ty = {JOUR},
	Url = {http://dx.doi.org/10.1038/nature23268},
	Volume = {547},
	Year = {2017}}

@article{Po2017,
	Author = {Po, Hoi Chun and Vishwanath, Ashvin and Watanabe, Haruki},
	Da = {2017/06/30},
	Doi = {10.1038/s41467-017-00133-2},
	Id = {Po2017},
	Isbn = {2041-1723},
	Journal = {Nature Communications},
	Number = {1},
	Pages = {50},
	title = {{Symmetry-based indicators of band topology in the 230 space groups}},
	Ty = {JOUR},
	Url = {https://doi.org/10.1038/s41467-017-00133-2},
	Volume = {8},
	Year = {2017}}

@article{Bouhon2020b,
  title = {{Geometric approach to fragile topology beyond symmetry indicators}},
  author = {Bouhon, Adrien and Bzdusek, Tom\'a\v{s} and Slager, {Robert-Jan}},
  journal = {Phys. Rev. B},
  volume = {102},
  issue = {11},
  pages = {115135},
  numpages = {29},
  year = {2020},
  month = {Sep},
  publisher = {American Physical Society},
  doi = {10.1103/PhysRevB.102.115135},
  url = {https://link.aps.org/doi/10.1103/PhysRevB.102.115135}
}

@misc{Bouhon2023,
      title = {{Quantum geometry beyond projective single bands}}, 
      author={Adrien Bouhon and Abigail Timmel and {Robert-Jan} Slager},
      year={2023},
      eprint={2303.02180},
      archivePrefix={arXiv},
      primaryClass={cond-mat.mes-hall},
      url={https://arxiv.org/abs/2303.02180}, 
}

@article{Provost1980,
  title = {{Riemannian structure on manifolds of quantum states}},
  author={Provost, JP and Vallee, G},
  journal={Communications in Mathematical Physics},
  volume={76},
  number={3},
  pages={289--301},
  year={1980},
  publisher={Springer},
  doi = {10.1007/BF02193559}
}

@article{Resta_2011,
   title = {{The insulating state of matter: a geometrical theory}},
   volume={79},
   ISSN={1434-6036},
   url={http://dx.doi.org/10.1140/epjb/e2010-10874-4},
   DOI={10.1140/epjb/e2010-10874-4},
   number={2},
   journal={The European Physical Journal B},
   publisher={Springer Science and Business Media LLC},
   author={Resta, R.},
   year={2011},
   month=jan, pages={121–137} }

@article{Brouwer2023,
  title = {{Homotopic classification of band structures: Stable, fragile, delicate, and stable representation-protected topology}},
  author = {Brouwer, Piet W. and Dwivedi, Vatsal},
  journal = {Phys. Rev. B},
  volume = {108},
  issue = {15},
  pages = {155137},
  numpages = {41},
  year = {2023},
  month = {Oct},
  publisher = {American Physical Society},
  doi = {10.1103/PhysRevB.108.155137},
  url = {https://link.aps.org/doi/10.1103/PhysRevB.108.155137}
}

@article{Yu2025qgt,
	author = {Yu, Jiabin and Bernevig, B. Andrei and Queiroz, Raquel and Rossi, Enrico and T{\"o}rm{\"a}, P{\"a}ivi and Yang, Bohm-Jung},
	date = {2025/10/10},
	doi = {10.1038/s41535-025-00801-3},
	id = {Yu2025},
	isbn = {2397-4648},
	journal = {npj Quantum Materials},
	number = {1},
	pages = {101},
	title = {Quantum geometry in quantum materials},
	url = {https://doi.org/10.1038/s41535-025-00801-3},
	volume = {10},
	year = {2025}}

@article{Ahn2020,
  title = {{Low-Frequency Divergence} and {Quantum Geometry} of the {Bulk Photovoltaic Effect} in {Topological Semimetals}},
  author = {Ahn, Junyeong and Guo, Guang-Yu and Nagaosa, Naoto},
  journal = {Phys. Rev. X},
  volume = {10},
  issue = {4},
  pages = {041041},
  numpages = {28},
  year = {2020},
  month = {Nov},
  publisher = {American Physical Society},
  doi = {10.1103/PhysRevX.10.041041},
  url = {https://link.aps.org/doi/10.1103/PhysRevX.10.041041}
}

@article{Torma2023,
  title = {{Essay: Where Can Quantum Geometry Lead Us?}},
  author = {T\"orm\"a, P\"aivi},
  journal = {Phys. Rev. Lett.},
  volume = {131},
  issue = {24},
  pages = {240001},
  numpages = {7},
  year = {2023},
  month = {Dec},
  publisher = {American Physical Society},
  doi = {10.1103/PhysRevLett.131.240001},
  url = {https://link.aps.org/doi/10.1103/PhysRevLett.131.240001}
}

@article{Hatano1996,
  title = {{Localization Transitions in Non-Hermitian Quantum Mechanics}},
  author = {Hatano, Naomichi and Nelson, David R.},
  journal = {Phys. Rev. Lett.},
  volume = {77},
  issue = {3},
  pages = {570--573},
  numpages = {0},
  year = {1996},
  month = {Jul},
  publisher = {American Physical Society},
  doi = {10.1103/PhysRevLett.77.570},
  url = {https://link.aps.org/doi/10.1103/PhysRevLett.77.570}
}

@article{Bender_2007,
doi = {10.1088/0034-4885/70/6/R03},
url = {https://doi.org/10.1088/0034-4885/70/6/R03},
year = {2007},
month = {may},
publisher = {},
volume = {70},
number = {6},
pages = {947},
author = {Bender, Carl M},
title = {Making sense of non-Hermitian {Hamiltonians}},
journal = {Reports on Progress in Physics}
}

@article{Kunst2021,
  title = {Exceptional topology of non-{Hermitian} systems},
  author = {Bergholtz, Emil J. and Budich, Jan Carl and Kunst, Flore K.},
  journal = {Rev. Mod. Phys.},
  volume = {93},
  issue = {1},
  pages = {015005},
  numpages = {31},
  year = {2021},
  month = {Feb},
  publisher = {American Physical Society},
  doi = {10.1103/RevModPhys.93.015005},
  url = {https://link.aps.org/doi/10.1103/RevModPhys.93.015005}
}

@article{Borgnia2020,
  title = {{Non-Hermitian Boundary Modes} and {Topology}},
  author = {Borgnia, Dan S. and Kruchkov, Alex Jura and Slager, Robert-Jan},
  journal = {Phys. Rev. Lett.},
  volume = {124},
  issue = {5},
  pages = {056802},
  numpages = {6},
  year = {2020},
  month = {Feb},
  publisher = {American Physical Society},
  doi = {10.1103/PhysRevLett.124.056802},
  url = {https://link.aps.org/doi/10.1103/PhysRevLett.124.056802}
}

@article{Zhou2019,
  title = {Periodic table for topological bands with non-{Hermitian} symmetries},
  author = {Zhou, Hengyun and Lee, Jong Yeon},
  journal = {Phys. Rev. B},
  volume = {99},
  issue = {23},
  pages = {235112},
  numpages = {19},
  year = {2019},
  month = {Jun},
  publisher = {American Physical Society},
  doi = {10.1103/PhysRevB.99.235112},
  url = {https://link.aps.org/doi/10.1103/PhysRevB.99.235112}
}

@article{Kawabata2019,
  title = {{Symmetry} and {Topology} in {Non-Hermitian Physics}},
  author = {Kawabata, Kohei and Shiozaki, Ken and Ueda, Masahito and Sato, Masatoshi},
  journal = {Phys. Rev. X},
  volume = {9},
  issue = {4},
  pages = {041015},
  numpages = {52},
  year = {2019},
  month = {Oct},
  publisher = {American Physical Society},
  doi = {10.1103/PhysRevX.9.041015},
  url = {https://link.aps.org/doi/10.1103/PhysRevX.9.041015}
}

@article{Okuma2020,
  title = {{Topological Origin of Non-Hermitian Skin Effects}},
  author = {Okuma, Nobuyuki and Kawabata, Kohei and Shiozaki, Ken and Sato, Masatoshi},
  journal = {Phys. Rev. Lett.},
  volume = {124},
  issue = {8},
  pages = {086801},
  numpages = {7},
  year = {2020},
  month = {Feb},
  publisher = {American Physical Society},
  doi = {10.1103/PhysRevLett.124.086801},
  url = {https://link.aps.org/doi/10.1103/PhysRevLett.124.086801}
}

@article{Loring2015,
   title={{K}-{Theory} and {Pseudospectra} for {Topological Insulators}},
   volume={356},
   ISSN={0003-4916},
   url={http://dx.doi.org/10.1016/j.aop.2015.02.031},
   DOI={10.1016/j.aop.2015.02.031},
   journal={Annals of Physics},
   publisher={Elsevier BV},
   author={Loring, Terry A.},
   year={2015},
   month=may, pages={383–416} }

@misc{Loring2017,
      title={Finite volume calculation of {$K$}-theory invariants}, 
      author={Terry Loring and Hermann Schulz-Baldes},
      year={2017},
      eprint={1701.07455},
      archivePrefix={arXiv},
      primaryClass={math-ph},
      url={https://arxiv.org/abs/1701.07455}, 
}

@misc{Loring2019,
      title={{A Guide to the Bott Index and Localizer Index}}, 
      author={Terry A. Loring},
      year={2019},
      eprint={1907.11791},
      archivePrefix={arXiv},
      primaryClass={math-ph},
      url={https://arxiv.org/abs/1907.11791}, 
}

@article{Cerjan2024,
   title={Classifying photonic topology using the spectral localizer and numerical {K}-theory},
   volume={9},
   ISSN={2378-0967},
   url={http://dx.doi.org/10.1063/5.0239018},
   DOI={10.1063/5.0239018},
   number={11},
   journal={APL Photonics},
   publisher={AIP Publishing},
   author={Cerjan, Alexander and Loring, Terry A.},
   year={2024},
   month=nov }

@article{Chadha2024,
  title = {Real-space topological localizer index to fully characterize the dislocation skin effect},
  author = {Chadha, Nisarg and Moghaddam, Ali G. and van den Brink, Jeroen and Fulga, Cosma},
  journal = {Phys. Rev. B},
  volume = {109},
  issue = {3},
  pages = {035425},
  numpages = {8},
  year = {2024},
  month = {Jan},
  publisher = {American Physical Society},
  doi = {10.1103/PhysRevB.109.035425},
  url = {https://link.aps.org/doi/10.1103/PhysRevB.109.035425}
}

@article{Jezequel2025,
	title = {{Explicit equivalence between the spectral localizer and local Chern and winding markers}},
	pages = {118},
	author = {Jezequel, Lucien and Bardarson, Jens H. and Grushin, Adolfo G.},
	journal = {SciPost Phys.},
	volume = {20},
	year = {2026},
	publisher = {SciPost},
	doi = {10.21468/SciPostPhys.20.4.118},
	url = {https://scipost.org/10.21468/SciPostPhys.20.4.118}
}

@article{Prosen2008,
doi = {10.1088/1367-2630/10/4/043026},
url = {https://doi.org/10.1088/1367-2630/10/4/043026},
year = {2008},
month = {apr},
publisher = {},
volume = {10},
number = {4},
pages = {043026},
author = {Prosen, Tomaž},
title = {Third quantization: a general method to solve master equations for quadratic open {Fermi} systems},
journal = {New Journal of Physics}
}

@article{Prosen2010,
doi = {10.1088/1742-5468/2010/07/P07020},
url = {https://doi.org/10.1088/1742-5468/2010/07/P07020},
year = {2010},
month = {jul},
publisher = {},
volume = {2010},
number = {07},
pages = {P07020},
author = {Prosen, Tomaž},
title = {{Spectral} theorem for the {Lindblad} equation for quadratic open fermionic systems},
journal = {Journal of Statistical Mechanics: Theory and Experiment}
}

@article{Bardyn2013,
doi = {10.1088/1367-2630/15/8/085001},
url = {https://doi.org/10.1088/1367-2630/15/8/085001},
year = {2013},
month = {aug},
publisher = {IOP Publishing},
volume = {15},
number = {8},
pages = {085001},
author = {Bardyn, C-E and Baranov, M A and Kraus, C V and Rico, E and İmamoğlu, A and Zoller, P and Diehl, S},
title = {Topology by dissipation},
journal = {New Journal of Physics}
}

@article{Lieu2020,
  title = {{Tenfold Way} for {Quadratic Lindbladians}},
  author = {Lieu, Simon and McGinley, Max and Cooper, Nigel R.},
  journal = {Phys. Rev. Lett.},
  volume = {124},
  issue = {4},
  pages = {040401},
  numpages = {6},
  year = {2020},
  month = {Jan},
  publisher = {American Physical Society},
  doi = {10.1103/PhysRevLett.124.040401},
  url = {https://link.aps.org/doi/10.1103/PhysRevLett.124.040401}
}

@article{Diehl2011,
	author = {Diehl, Sebastian and Rico, Enrique and Baranov, Mikhail A. and Zoller, Peter},
	date = {2011/12/01},
	doi = {10.1038/nphys2106},
	id = {Diehl2011},
	isbn = {1745-2481},
	journal = {Nature Physics},
	number = {12},
	pages = {971--977},
	title = {Topology by dissipation in atomic quantum wires},
	url = {https://doi.org/10.1038/nphys2106},
	volume = {7},
	year = {2011}}

@article{Zhou2022,
  title = {Non-Hermitian skin effect in quadratic {Lindbladian} systems: {An} adjoint fermion approach},
  author = {Zhou, Ziheng and Yu, Zhenhua},
  journal = {Phys. Rev. A},
  volume = {106},
  issue = {3},
  pages = {032216},
  numpages = {7},
  year = {2022},
  month = {Sep},
  publisher = {American Physical Society},
  doi = {10.1103/PhysRevA.106.032216},
  url = {https://link.aps.org/doi/10.1103/PhysRevA.106.032216}
}

@article{Yang2022,
  title = {Liouvillian skin effect in an exactly solvable model},
  author = {Yang, Fan and Jiang, Qing-Dong and Bergholtz, Emil J.},
  journal = {Phys. Rev. Res.},
  volume = {4},
  issue = {2},
  pages = {023160},
  numpages = {19},
  year = {2022},
  month = {May},
  publisher = {American Physical Society},
  doi = {10.1103/PhysRevResearch.4.023160},
  url = {https://link.aps.org/doi/10.1103/PhysRevResearch.4.023160}
}

@article{Haga2021,
  title = {{Liouvillian Skin Effect}: {Slowing Down} of {Relaxation Processes} without {Gap Closing}},
  author = {Haga, Taiki and Nakagawa, Masaya and Hamazaki, Ryusuke and Ueda, Masahito},
  journal = {Phys. Rev. Lett.},
  volume = {127},
  issue = {7},
  pages = {070402},
  numpages = {7},
  year = {2021},
  month = {Aug},
  publisher = {American Physical Society},
  doi = {10.1103/PhysRevLett.127.070402},
  url = {https://link.aps.org/doi/10.1103/PhysRevLett.127.070402}
}

@article{Minganti2020,
  title = {{Hybrid-Liouvillian} formalism connecting exceptional points of non-{Hermitian Hamiltonians} and {Liouvillians} via postselection of quantum trajectories},
  author = {Minganti, Fabrizio and Miranowicz, Adam and Chhajlany, Ravindra W. and Arkhipov, Ievgen I. and Nori, Franco},
  journal = {Phys. Rev. A},
  volume = {101},
  issue = {6},
  pages = {062112},
  numpages = {14},
  year = {2020},
  month = {Jun},
  publisher = {American Physical Society},
  doi = {10.1103/PhysRevA.101.062112},
  url = {https://link.aps.org/doi/10.1103/PhysRevA.101.062112}
}

@article{Michishita2020,
  title = {{Equivalence} of {Effective Non-Hermitian Hamiltonians} in the {Context} of {Open Quantum Systems} and {Strongly Correlated Electron Systems}},
  author = {Michishita, Yoshihiro and Peters, Robert},
  journal = {Phys. Rev. Lett.},
  volume = {124},
  issue = {19},
  pages = {196401},
  numpages = {6},
  year = {2020},
  month = {May},
  publisher = {American Physical Society},
  doi = {10.1103/PhysRevLett.124.196401},
  url = {https://link.aps.org/doi/10.1103/PhysRevLett.124.196401}
}

@article{Wanjura2020,
	author = {Wanjura, Clara C. and Brunelli, Matteo and Nunnenkamp, Andreas},
	date = {2020/06/19},
	doi = {10.1038/s41467-020-16863-9},
	id = {Wanjura2020},
	isbn = {2041-1723},
	journal = {Nature Communications},
	number = {1},
	pages = {3149},
	title = {Topological framework for directional amplification in driven-dissipative cavity arrays},
	url = {https://doi.org/10.1038/s41467-020-16863-9},
	volume = {11},
	year = {2020}}

@article{Chaduteau2026,
  title = {{Lindbladian} versus {Postselected} non-{Hermitian} {Topology}},
  author = {Chaduteau, Alexandre and Lee, Derek K. K. and Schindler, Frank},
  journal = {Phys. Rev. Lett.},
  volume = {136},
  issue = {1},
  pages = {016603},
  numpages = {6},
  year = {2026},
  month = {Jan},
  publisher = {American Physical Society},
  doi = {10.1103/ljvt-w6hw},
  url = {https://link.aps.org/doi/10.1103/ljvt-w6hw}
}

@misc{Nemeth2026,
      title={{Operator Space Transport} and the {Emergence of Boundary Time Crystals}}, 
      author={Dominik Nemeth and Ahsan Nazir and Robert-Jan Slager and Alessandro Principi},
      year={2026},
      eprint={2604.14291},
      archivePrefix={arXiv},
      primaryClass={quant-ph},
      url={https://arxiv.org/abs/2604.14291}, 
}

@article{Byrd2003,
  title = {Characterization of the positivity of the density matrix in terms of the coherence vector representation},
  author = {Byrd, Mark S. and Khaneja, Navin},
  journal = {Phys. Rev. A},
  volume = {68},
  issue = {6},
  pages = {062322},
  numpages = {13},
  year = {2003},
  month = {Dec},
  publisher = {American Physical Society},
  doi = {10.1103/PhysRevA.68.062322},
  url = {https://link.aps.org/doi/10.1103/PhysRevA.68.062322}
}

@article{Kimura2003,
title = {{The Bloch} vector for {N}-level systems},
journal = {Physics Letters A},
volume = {314},
number = {5},
pages = {339-349},
year = {2003},
issn = {0375-9601},
doi = {https://doi.org/10.1016/S0375-9601(03)00941-1},
url = {https://www.sciencedirect.com/science/article/pii/S0375960103009411},
author = {Gen Kimura}
}

@article{Bertlmann2008,
doi = {10.1088/1751-8113/41/23/235303},
url = {https://doi.org/10.1088/1751-8113/41/23/235303},
year = {2008},
month = {may},
publisher = {},
volume = {41},
number = {23},
pages = {235303},
author = {Bertlmann, Reinhold A and Krammer, Philipp},
title = {Bloch vectors for qudits},
journal = {Journal of Physics A: Mathematical and Theoretical}
}

@article{Byrd2011,
  title = {General open-system quantum evolution in terms of affine maps of the polarization vector},
  author = {Byrd, Mark S. and Bishop, C. Allen and Ou, Yong-Cheng},
  journal = {Phys. Rev. A},
  volume = {83},
  issue = {1},
  pages = {012301},
  numpages = {13},
  year = {2011},
  month = {Jan},
  publisher = {American Physical Society},
  doi = {10.1103/PhysRevA.83.012301},
  url = {https://link.aps.org/doi/10.1103/PhysRevA.83.012301}
}

@misc{Lambert2024,
      title={{QuTiP} 5: The {Quantum Toolbox} in {Python}}, 
      author={Neill Lambert and Eric Giguère and Paul Menczel and Boxi Li and Patrick Hopf and Gerardo Suárez and Marc Gali and Jake Lishman and Rushiraj Gadhvi and Rochisha Agarwal and Asier Galicia and Nathan Shammah and Paul Nation and J. R. Johansson and Shahnawaz Ahmed and Simon Cross and Alexander Pitchford and Franco Nori},
      year={2024},
      eprint={2412.04705},
      archivePrefix={arXiv},
      primaryClass={quant-ph},
      url={https://arxiv.org/abs/2412.04705}, 
}

@article{Fano1957,
  title = {{Description} of {States} in {Quantum Mechanics} by {Density Matrix} and {Operator Techniques}},
  author = {Fano, U.},
  journal = {Rev. Mod. Phys.},
  volume = {29},
  issue = {1},
  pages = {74--93},
  numpages = {0},
  year = {1957},
  month = {Jan},
  publisher = {American Physical Society},
  doi = {10.1103/RevModPhys.29.74},
  url = {https://link.aps.org/doi/10.1103/RevModPhys.29.74}
}

@article{Happer1972,
  title = {{Optical} {Pumping}},
  author = {Happer, William},
  journal = {Rev. Mod. Phys.},
  volume = {44},
  issue = {2},
  pages = {169--249},
  numpages = {0},
  year = {1972},
  month = {Apr},
  publisher = {American Physical Society},
  doi = {10.1103/RevModPhys.44.169},
  url = {https://link.aps.org/doi/10.1103/RevModPhys.44.169}
}

@article{Wood2014,
  title = {Cavity Cooling of an Ensemble Spin System},
  author = {Wood, Christopher J. and Borneman, Troy W. and Cory, David G.},
  journal = {Phys. Rev. Lett.},
  volume = {112},
  issue = {5},
  pages = {050501},
  numpages = {5},
  year = {2014},
  month = {Feb},
  publisher = {American Physical Society},
  doi = {10.1103/PhysRevLett.112.050501},
  url = {https://link.aps.org/doi/10.1103/PhysRevLett.112.050501}
}

@Inbook{Blum2012,
author="Blum, Karl",
title="{Irreducible Components} of the {Density Matrix}",
bookTitle="Density Matrix Theory and Applications",
year="2012",
publisher="Springer Berlin Heidelberg",
address="Berlin, Heidelberg",
pages="115--163",
isbn="978-3-642-20561-3",
doi="10.1007/978-3-642-20561-3_4",
url="https://doi.org/10.1007/978-3-642-20561-3_4"
}

@Inbook{Hecht2000,
author="Hecht, K. T.",
title="{{Spherical Tensor Operators}}",
bookTitle="Quantum Mechanics",
year="2000",
publisher="Springer New York",
address="New York, NY",
pages="294--298",
isbn="978-1-4612-1272-0",
doi="10.1007/978-1-4612-1272-0_31",
url="https://doi.org/10.1007/978-1-4612-1272-0_31"
}

@book{Varshalovich1988,
  address = {Singapore},
  author = {Varshalovich, D. A. and Moskalev, A. N. and Khersonskii, V. K.},
  publisher = {World Scientific},
  title = {{Quantum Theory} of {Angular Momentum}},
  year = 1988
}

@article{Walls1978,
	author = {Walls, D. F. and Drummond, P. D. and Hassan, S. S. and Carmichael, H. J.},
	doi = {10.1143/PTPS.64.307},
	issn = {0375-9687},
	journal = {Progress of Theoretical Physics Supplement},
	month = {02},
	pages = {307-320},
	title = {{Non-Equilibrium Phase Transitions} in {Cooperative Atomic Systems}},
	url = {https://doi.org/10.1143/PTPS.64.307},
	volume = {64},
	year = {1978}}

@article{Drummond1978,
title = {Volterra cycles and the cooperative fluorescence critical point},
journal = {Optics Communications},
volume = {27},
number = {1},
pages = {160-164},
year = {1978},
issn = {0030-4018},
doi = {https://doi.org/10.1016/0030-4018(78)90198-0},
url = {https://www.sciencedirect.com/science/article/pii/0030401878901980},
author = {P.D. Drummond and H.J. Carmichael}
}

@article{Drummond1980,
  title = {Observables and moments of cooperative resonance fluorescence},
  author = {Drummond, P. D.},
  journal = {Phys. Rev. A},
  volume = {22},
  issue = {3},
  pages = {1179--1184},
  numpages = {0},
  year = {1980},
  month = {Sep},
  publisher = {American Physical Society},
  doi = {10.1103/PhysRevA.22.1179},
  url = {https://link.aps.org/doi/10.1103/PhysRevA.22.1179}
}

@article{Schneider2002,
  title = {{Entanglement in the steady state of a collective-angular-momentum (Dicke) model}},
  author = {Schneider, S. and Milburn, G. J.},
  journal = {Phys. Rev. A},
  volume = {65},
  issue = {4},
  pages = {042107},
  numpages = {5},
  year = {2002},
  month = {Mar},
  publisher = {American Physical Society},
  doi = {10.1103/PhysRevA.65.042107},
  url = {https://link.aps.org/doi/10.1103/PhysRevA.65.042107}
}

@article{Hannukainen2018,
  title = {Dissipation-driven quantum phase transitions and symmetry breaking},
  author = {Hannukainen, Julia and Larson, Jonas},
  journal = {Phys. Rev. A},
  volume = {98},
  issue = {4},
  pages = {042113},
  numpages = {11},
  year = {2018},
  month = {Oct},
  publisher = {American Physical Society},
  doi = {10.1103/PhysRevA.98.042113},
  url = {https://link.aps.org/doi/10.1103/PhysRevA.98.042113}
}

@article{Ribeiro2019,
  title = {{Integrable Quantum Dynamics} of {Open Collective Spin Models}},
  author = {Ribeiro, Pedro and Prosen, Tomaž},
  journal = {Phys. Rev. Lett.},
  volume = {122},
  issue = {1},
  pages = {010401},
  numpages = {6},
  year = {2019},
  month = {Jan},
  publisher = {American Physical Society},
  doi = {10.1103/PhysRevLett.122.010401},
  url = {https://link.aps.org/doi/10.1103/PhysRevLett.122.010401}
}

@incollection{Breuer2007,
	author = {Breuer, Heinz-Peter and Petruccione, Francesco},
	booktitle = {{The Theory of Open Quantum Systems}},
	doi = {10.1093/acprof:oso/9780199213900.003.03},
	isbn = {9780199213900},
	month = {01},
	publisher = {Oxford University Press},
	title = {Quantum Master Equations},
	url = {https://doi.org/10.1093/acprof:oso/9780199213900.003.03},
	year = {2007}}

@article{Nakagawa2025,
  title = {{Topology of Discrete Quantum Feedback Control}},
  author = {Nakagawa, Masaya and Ueda, Masahito},
  journal = {Phys. Rev. X},
  volume = {15},
  issue = {2},
  pages = {021016},
  numpages = {46},
  year = {2025},
  month = {Apr},
  publisher = {American Physical Society},
  doi = {10.1103/PhysRevX.15.021016},
  url = {https://link.aps.org/doi/10.1103/PhysRevX.15.021016}
}

@misc{Koch2025,
      title={{Liouvillian topology and non-reciprocal dynamics in open Floquet chains}}, 
      author={Florian Koch and Yu-Min Hu and Jan Carl Budich},
      year={2025},
      eprint={2511.16742},
      archivePrefix={arXiv},
      primaryClass={cond-mat.mes-hall},
      url={https://arxiv.org/abs/2511.16742}, 
}

@misc{Nakagawa2026,
      title={{Topological Characterization of Discrete-Time Classical Stochastic Processes: Dual Role of Point-Gap Topology}}, 
      author={Masaya Nakagawa and Masahito Ueda},
      year={2026},
      eprint={2605.07231},
      archivePrefix={arXiv},
      primaryClass={cond-mat.mes-hall},
      url={https://arxiv.org/abs/2605.07231}, 
}

@misc{nemeth2026code,
  author       = {Dominik Nemeth},
 title        = {Python code: {Topological Boundary Time Crystal Oscillations}},
  year         = {2026},
  howpublished = {\url{https://github.com/d-nemeth/spectral-localizer-boundary-time-crystal}}
}

@article{Gyamfi2020,
doi = {10.1088/1361-6404/ab9fdd},
url = {https://doi.org/10.1088/1361-6404/ab9fdd},
year = {2020},
month = {oct},
publisher = {IOP Publishing},
volume = {41},
number = {6},
pages = {063002},
author = {Gyamfi, Jerryman A},
title = {Fundamentals of quantum mechanics in {Liouville} space},
journal = {European Journal of Physics}
}

\appendix

\section*{Notation and Conventions}
\label{sec:notation_conventions}
In the BTC literature, it is standard to work within the maximally symmetric sector, $j=N/2$, and we adopt this convention throughout.

%Nevertheless, the formalism developed here applies generally to arbitrary spin-$j$ representations upon identifying $N=2j$. More generally, our arguments rely only on the existence of a large-spin collective limit, $j\to\infty$, under which the decay rates in the BTC phase become vanishingly small.

Throughout the Appendices calligraphic symbols ($\mathcal A$) are used for superoperators acting on the Hilbert space of operators $\mathcal B (\mathcal H_j)$, whereas double-lined symbols ($\mathbb A$) are used for superoperators acting on the vectorized, Liouville space $\mathcal H_j \otimes \mathcal H_j^*$. In addition, the vectorized form of an operator $O$ is denoted via $\ket{O}\rangle$. For more detail, we refer the Reader to Ref.~\cite{Gyamfi2020}.

\section{Details on Spectral Localizer}
We prove that the definition of the spectral localizer in Eq.~(\ref{eq:spectral_localizer}) is equivalent, up to a Pauli basis rotation, to the definition $L = (\Tilde{X} + i \Tilde{H})\Gamma$ as is typically used in NH, one-dimensional chains \cite{Chadha2024}. Here, we set $\kappa=1$, $\lambda_0=0$ and $x_0=0$ for ease of notation. The Hermitian Hamiltonian
\begin{equation}
    \Tilde{H} = 
    \begin{pmatrix}
        0 & \mathcal L \\
        \mathcal L^\dagger & 0 \\ 
    \end{pmatrix},
\end{equation}
is in class AIII~\cite{Chadha2024}, $\Tilde{X}=\mathrm{diag}(\mathcal X, \mathcal X)$ and $\Gamma=\mathrm{diag}(\mathbb I, - \mathbb I)$ is a chiral matrix, such that $\Tilde{H}$ exhibits the chiral symmetry $\Gamma \Tilde{H} = - \Tilde{H} \Gamma$. Using these definitions, one can show that
\begin{equation}
    L =
    \begin{pmatrix}
        \mathcal X & -i\mathcal L\\
        i \mathcal L^\dagger & - \mathcal X
    \end{pmatrix},
\end{equation}
which by writing $\mathcal L = \mathrm{Re}(\mathcal L) + i \, \mathrm{Im}(\mathcal L) $, where $\mathrm{Re}(\mathcal L) =\tfrac{1}{2}(\mathcal L + \mathcal L^\dagger)$ and $\mathrm{Im}(\mathcal L) =\tfrac{1}{2i}(\mathcal L - \mathcal L^\dagger)$ results in
\begin{equation}
    L = \mathrm{Re}(\mathcal{L})\otimes \sigma_y + \mathrm{Im}(\mathcal{L})\otimes \sigma_x + \mathcal X \otimes \sigma_z.
\end{equation}
This is equivalent to Eq.~(\ref{eq:spectral_localizer}) up to a rotation in of the Pauli basis $\sigma_{x,y,z}$. Therefore, the indices $\nu^L$ obtained through these are equal in magnitude and differ only in their sign. We employ our definition as in Eq.~(\ref{eq:spectral_localizer}) to better motivate the meaning behind the local Chern-like markers in spectral space.

Note that this construction should not be confused with genuine Chern markers, which arise when the underlying base space is two-dimensional in physical or effective spatial coordinates. In the present case, the index is a point-gap invariant evaluated locally as a function of the complex spectral parameter $\lambda_0$. The resulting two-dimensional structure in the $\lambda$-plane reflects a complex frequency-resolved obstruction to simultaneous spectral and operator space localization, rather than a Chern curvature defined over a spatial manifold. The $\lambda$-plane plays the role of a parameter space over which $\nu^L$ is evaluated, producing isolated topological “islands” in spectral space.

\section{Mapping to a NH Hopping Model}
\label{sec:mapping_nh_hopping_model}
In this section we briefly outline how the BTC model [Eq.~(\ref{eq:BTC_master_Eq})] can be mapped onto an effective NH hopping model [Eq.~(\ref{eq:non_hermitian_hopping_mapping})]. For a more complete discussion of this mapping and its physical implications we refer the Reader to Ref.~\cite{Nemeth2026}.

We begin by introducing the adjoint generators,
\begin{equation}
    \mathcal K_\alpha[A]
    :=
    [J_\alpha,A],
    \qquad
    \alpha=x,y,z,
\end{equation}
where $J_\alpha$ are the collective spin operators introduced in the main text and $A$ is an arbitrary operator. Whereas the operators $J_\alpha$ generate rotations while acting on states, the adjoint generators $\mathcal K_\alpha$ generate rotations while acting on operators. Since these generators correspond to the adjoint action of the $\mathrm{SU}(2)$ Lie group, they satisfy the same $\mathfrak{su}(2)$ algebra,
\begin{equation}
    [\mathcal K_\alpha,\mathcal K_\beta]
    =
    i\epsilon_{\alpha\beta\gamma}\mathcal K_\gamma.
\end{equation}
In direct analogy with the ordinary spin Casimir operator $J^2=J_x^2+J_y^2+J_z^2$, one may also define the adjoint Casimir operator,
\begin{equation}
\begin{aligned}
    \mathcal K^2[A] := \sum_{\alpha=x,y,z} \mathcal K_\alpha^2[A]
    &= \sum_{\alpha=x,y,z} [J_\alpha,[J_\alpha,A]].
\end{aligned}
\end{equation}

The spherical tensor operators $T^k_q$ are simultaneous eigenoperators of both $\mathcal K^2$ and $\mathcal K_z$~\cite{Varshalovich1988,Hecht2000},
\begin{equation}
\begin{aligned}
    \mathcal K^2[T^k_q] &= k(k+1)\,T^k_q, \\ \mathcal K_z[T^k_q] &= q\, T^k_q,
\end{aligned}
\end{equation}
with corresponding eigenvalues $k(k+1)$ and $q$, respectively. Furthermore, defining the raising and lowering superoperators
\begin{equation}
    \mathcal K_\pm
    :=
    \mathcal K_x\pm i\mathcal K_y,
\end{equation}
one obtains
\begin{equation}
    \mathcal K_\pm[T^k_q]
    =
    \sqrt{k(k+1)-q(q\pm1)}
    \,T^k_{q\pm1}.
\end{equation}
The set $\{T^k_q\}$ forms an orthonormal operator basis with respect to the Hilbert--Schmidt inner product~\cite{Blum2012},
\begin{equation}
\mathrm{Tr}\!\left[
(T^k_q)^\dagger
T^{k'}_{q'}
\right]
=
\delta_{kk'}\delta_{qq'}.
\label{eq:spherical_tensor_orthonormality}
\end{equation}

The Liouvillian superoperator in Eq.~(\ref{eq:BTC_master_Eq}) can be decomposed as
\begin{equation}
    \mathcal L
    =
    \mathcal L_0+\mathcal L_D,
\end{equation}
where
\begin{equation}
    \mathcal L_0[\rho]
    =
    -i\Omega[J_x,\rho]
\end{equation}
describes the coherent unitary evolution, while
\begin{equation}
    \mathcal L_D[\rho]
    =
    \frac{\Gamma}{N}
    \left(
    J_-\rho J_+
    -
    \frac12\{J_+J_-,\rho\}
    \right)
\end{equation}
describes the dissipative contribution.

We now analyze these two terms separately. Firstly, the coherent contribution may be rewritten in terms of the adjoint generators as
\begin{equation}
    \mathcal L_0
    =
    -\frac{i\Omega}{2}
    (\mathcal K_++\mathcal K_-).
\end{equation}
Since $\mathcal K_\pm$ map $T^k_q\to T^k_{q\pm1}$, the coherent contribution generates nearest-neighbor hopping in the $q$-direction with amplitudes
\begin{equation}
    -\frac{i\Omega}{2}
    \sqrt{k(k+1)-q(q\pm1)}.
\end{equation}
Thus, $\mathcal L_0$ generates coherent transport along the $q$-coordinate of operator space.

To analyze the dissipative contribution $\mathcal L_D$, we make use of a composition rule for irreducible tensors. For two irreducible tensors $V^{k_1}_{q_1}$ and $U^{k_2}_{q_2}$, one has~\cite{Varshalovich1988}
\begin{equation}
    V^{k_1}_{q_1}U^{k_2}_{q_2}
    =
    \sum_{k=\vert k_1-k_2\vert }^{k_1+k_2}
    \langle
    k_1q_1;
    k_2q_2
    \vert kq
    \rangle
    T^k_q,
\label{eq:direct_product_of_irreducible_tensors}
\end{equation}
where
\begin{equation}
\begin{aligned}
    k &= \vert k_1-k_2\vert , \vert k_1-k_2\vert +1, \ldots, k_1+k_2, \\ q&=q_1+q_2.
\end{aligned}
\end{equation}
Here,
$\langle k_1q_1;k_2q_2\vert kq\rangle$ denote Clebsch--Gordan coefficients.

Since $J_\pm$ are proportional to the rank-$1$ tensors $T^1_{\pm1}$, one can apply Eq.~(\ref{eq:direct_product_of_irreducible_tensors}) twice to determine the action of $\mathcal L_D$ on $T^k_q$. Multiplication by $J_-$ first generates tensors with
\begin{equation}
    k'=k-1,k,k+1,
    \qquad
    q'=q-1.
\end{equation}
Subsequent multiplication by $J_+$ produces terms with
\begin{equation}
\begin{aligned}
    k''&= k-2, k-1, k, k+1, k+2, \\ q''&=q.
\end{aligned}
\end{equation}
As a result, $\mathcal L_D$ generates short-range couplings in the $k$-direction together with on-site decay terms. The same method can be applied to the anti-commutator term. In fact, for the BTC model one can show that only couplings satisfying
\begin{equation}
    k\to k-1,k,k+1
\end{equation}
survive~\cite{Nemeth2026}. Consequently, the BTC Liouvillian maps onto an effective local NH hopping problem on the emergent $(k,q)$ operator space lattice, where $\mathcal L_0$ generates coherent hopping in $q$, while $\mathcal L_D$ generates dissipative hopping and decay in the $k$-direction.

This structure becomes particularly transparent in vectorized form, where the Lindbladian superoperator is represented by the matrix $\mathbb L$,
\begin{widetext}
\begin{equation}
\begin{aligned}
    \mathbb L ={}& -i\Omega \sum_{k,q}\Bigl[
        w_+(k,q-1)\vert k,q\rangle\rangle
        \langle\!\langle k,q-1\vert
 + w_-(k,q+1)\,\vert k,q\rangle\rangle
        \langle\!\langle k,q+1\vert\Bigr] \\
    &+ \frac{\Gamma}{N}\sum_{k,q}\Bigl[
        t_+(k-1,q)\,\vert k,q\rangle\rangle
        \langle\!\langle k-1,q\vert
 + t_-(k+1,q)\,\vert k,q\rangle\rangle
        \langle\!\langle k+1,q\vert 
 - \gamma(k,q)\,\vert k,q\rangle\rangle
        \langle\!\langle k,q\vert\Bigr],
\end{aligned}
\label{eq:vectorized_non_hermitian_hopping_model}
\end{equation}
\end{widetext}
where $w_\pm(k,q)$ and $t_\pm(k,q)$ are position dependent hopping amplitudes on the $(k,q)$ lattice and $\gamma(k,q)$ represents on-site decay.

Under vectorization, the master equation
\begin{equation}
    \frac{d}{dt}\rho(t)
    =
    \mathcal L[\rho(t)]
\end{equation}
is mapped to the linear equation
\begin{equation}
    \frac{d}{dt}
    \vert \rho(t)\rangle\rangle
    =
    \mathbb L
    \vert \rho(t)\rangle\rangle,
\end{equation}
where the vectorized density operator is expanded as
\begin{equation}
    \vert \rho(t)\rangle\rangle
    =
    \sum_{k'=0}^{2j}
    \sum_{q'=-k'}^{k'}
    a_{k',q'}(t)
    \vert k',q'\rangle\rangle.
    \label{eq:vectorized_spherical_tensor_expansion}
\end{equation}
Here, $\vert k,q\rangle\rangle$
denotes the vectorized spherical tensor operator corresponding to
\begin{equation}
    T^k_q
    \mapsto
    \vert k,q\rangle\rangle.
\end{equation}

Within this representation, the coherent contribution $\mathcal L_0$ generates reciprocal hopping in the $q$-direction, such that
\begin{equation}
    w_+(k,q)
    =
    w_-(k,q+1).
\end{equation}

In other words, the hopping amplitude for the process $q\to q+1$ is equal to that for the reverse process $q+1\to q$. By contrast, the dissipative contribution generates non-reciprocal hopping in the $k$-direction~\cite{Nemeth2026}, for which
\begin{equation}
    \vert t_+(k,q)\vert
  \neq
    \vert t_-(k+1,q)\vert,
\end{equation}
thereby inducing directional transport in operator space~\cite{Hatano1996, Borgnia2020, Okuma2020}. For more detail on the numerical values of $t_\pm(k,q)$ and $\gamma(k,q)$ we refer the Reader to Ref.~\cite{Nemeth2026}.

In the conventional setting of the NH skin effect~\cite{Hatano1996,Borgnia2020,Okuma2020}, the hopping amplitudes are spatially uniform and translation invariant, giving rise to a single global winding number. By contrast, the operator space hopping structure considered here is intrinsically position dependent and inhomogeneous. Consequently, the associated winding numbers depend on the local operator space coordinates and are therefore defined as local winding numbers.

For $\Gamma=0$, each eigenvector of $\mathbb{L}$ remains localized within a fixed-$k$ sector and can therefore be uniquely labelled by the quantum number $k$~\cite{Nemeth2026}. By contrast, for $\Gamma \neq 0$, the right eigenvectors acquire a net bias toward increasing $k$, reflecting the non-reciprocal transport induced by the dissipative dynamics. Correspondingly, the left eigenvectors, governed by $\mathbb{L}^\dagger$, acquire a net bias toward decreasing $k$.

Finally, the master equation $\frac{d}{dt}\rho= \mathcal L[\rho]$ can be expressed directly in terms of the expansion coefficients $\{a_{k,q}\}$ by substituting the vectorized representations from Eq.~(\ref{eq:vectorized_non_hermitian_hopping_model}) and Eq.~(\ref{eq:vectorized_spherical_tensor_expansion}) into the master equation. Making use of the orthonormality relations
\begin{equation}
    \langle\!\langle k,q|k',q'\rangle\!\rangle
    =
    \delta_{kk'}\delta_{qq'},
\end{equation}
which follow from Eq.~(\ref{eq:spherical_tensor_orthonormality}), one obtains
\begin{equation}
   \frac{d}{dt}
   a_{k,q}(t)
   =
   -i\Omega\,\mathcal W_{k,q}(t)
   +
   \frac{\Gamma}{2j}\,
   \mathcal T_{k,q}(t),
   \label{eq:hopping_model_coeffs_master_eqn}
\end{equation}
where
\begin{equation}
\begin{aligned}
    \mathcal W_{k,q}(t) &= w_+(k,q-1)\, a_{k,q-1}(t) \\ &\quad + w_-(k,q+1)\, a_{k,q+1}(t),
\end{aligned}
\end{equation}
and
\begin{equation}
\begin{aligned}
    \mathcal T_{k,q}(t) &= t_+(k-1,q)\, a_{k-1,q}(t) \\ &\quad + t_-(k+1,q)\, a_{k+1,q}(t) \\ &\quad - \gamma(k,q)\, a_{k,q}(t).
\end{aligned}
\end{equation}

Although Eq.~(\ref{eq:hopping_model_coeffs_master_eqn}) closely resembles a classical stochastic hopping equation~\cite{Breuer2007}, it remains a fully quantum Lindblad master equation expressed in the spherical tensor basis. 

%This distinction is important. In a conventional angular-momentum dyad expansion,
%\begin{equation}
%    \rho
%    =
%    \sum_{m,m'}
%    \rho_{mm'}
%    |j,m\rangle\langle j,m'|,
%\end{equation}
%restricting the dynamics to the diagonal sector $\rho_{mm}$ yields a classical rate equation for populations, where the diagonal matrix elements can be interpreted as probabilities. By contrast, the coefficients $a_{k,q}$ represent generally complex-valued operator amplitudes associated with irreducible multipole sectors of operator space.

%Consequently, the emergent hopping model should not be interpreted as a stochastic transport process for classical probabilities. Rather, it describes the coherent and dissipative transport of operator amplitudes in Liouville space, where the coordinates $(k,q)$ label collective spin operators.

The spherical tensor expansion utilized in this work is a direct generalization of the Bloch-vector representation of a qubit~\cite{Byrd2003,Kimura2003,Bertlmann2008,Byrd2011} to collective spin-$j$ systems.

%For a single qubit, the density operator may be written as
%\begin{equation}
%    \rho(t)
%    =
%    \frac12
%    \Bigl(
%    \mathbf1
%    +
%    \mathbf r(t)\cdot\boldsymbol{\sigma}
%    \Bigr),
%\end{equation}
%where
%$\boldsymbol{\sigma}=(\sigma_x,\sigma_y,\sigma_z)^T$
%denotes the Pauli operators and $\mathbf r(t)$ is the Bloch vector. The Bloch-vector components are directly related to observable expectation values through
%\begin{equation}
%    \langle \sigma_\alpha (t)\rangle = r_\alpha(t),
%\end{equation}
%where we have used the orthogonality relation
%\begin{equation}
%    \Tr(\sigma_\alpha^\dagger\sigma_\beta)
%    =
%    2\delta_{\alpha\beta}
%\end{equation}
%together with standard Bloch-vector identities~\cite{Byrd2003,Kimura2003,Bertlmann2008,Byrd2011}. Thus, the Bloch-vector components are related to the expectation values of the corresponding spin operators.

\section{Exact Equations of Motion}
\label{sec:exact_eom}
We consider the collective-spin Lindblad master equation describing the BTC model,
\begin{equation}
\begin{aligned}
    \frac{d}{dt}\rho &= \mathcal L[\rho] \\
    &= -i\Omega[J_x,\rho] \\
    &\quad + \frac{\Gamma}{N}\left(
        J_-\rho J_+ -\frac{1}{2}\{J_+J_-,\rho\}\right).
\end{aligned}
\end{equation}

The corresponding adjoint Liouvillian, governing the Heisenberg-picture evolution of observables is given by
\begin{equation}
\begin{aligned}
    \mathcal L^\dagger[O] &= i\Omega[J_x,O] \\
    &\quad + \frac{\Gamma}{2N}\left(
        J_+[O,J_-]+[J_+,O]J_-\right),
\end{aligned}
\end{equation}
which follows from expanding the anticommutator in the dissipator. Expectation values then evolve according to
\begin{equation}
\frac{d}{dt}\langle O\rangle
=
\left\langle \mathcal L^\dagger[O]\right\rangle.
\end{equation}
We now derive the exact equations of motion for the collective spin operators $J_\alpha$, $\alpha=x,y,z$.

\subsection*{Hamiltonian Contributions}

Using the angular momentum commutation relations $[J_\alpha,J_\beta]=i\epsilon_{\alpha\beta\gamma}J_\gamma$, one finds
\begin{equation}
\begin{split}
    \left.\frac{d}{dt}J_x\right\vert_H &= i\Omega[J_x,J_x]=0, \\
    \left.\frac{d}{dt}J_y\right\vert_H &= i\Omega[J_x,J_y]=-\Omega J_z, \\
    \left.\frac{d}{dt}J_z\right\vert_H &= i\Omega[J_x,J_z]=\Omega J_y.
\end{split}
\end{equation}

\subsection*{Dissipative Contributions}

Using the identities $J_\pm = J_x \pm i J_y$ together with the commutation relations, the dissipative contributions can be written in compact form as
\begin{equation}
\begin{split}
    \left.\frac{d}{dt}J_x\right\vert_D &= \frac{\Gamma}{2N}
    \left(
    \{J_x,J_z\}-J_x
    \right), \\
    \left.\frac{d}{dt}J_y\right\vert_D &= \frac{\Gamma}{2N}
    \left(
    \{J_y,J_z\}-J_y
    \right), \\
    \left.\frac{d}{dt}J_z\right\vert_D &= -\frac{\Gamma}{N}
    \left(
    J_x^2 + J_y^2 + J_z
    \right).
\end{split}
\end{equation}

Combining both contributions yields a closed set of equations for the first moments that explicitly depend on second-order operator moments.

\section{Mean-Field Approximation}
\label{sec:mf}
\subsection{Mean-Field Equations of Motion}

In semiclassical treatments one introduces the normalized magnetization variables
\begin{equation}
\begin{aligned}
    m_\alpha = \frac{2}{N}\langle J_\alpha \rangle = \frac{1}{j}\langle J_\alpha \rangle, \qquad j=\frac{N}{2},
\end{aligned}
\end{equation}
which become classical variables in the thermodynamic limit~\cite{Iemini2018, Piccito2021}.

The mean-field approximation assumes factorization of expectation values,
\begin{equation}
\langle J_\alpha J_\beta\rangle \approx \langle J_\alpha \rangle \langle J_\beta\rangle,
\end{equation}
which is justified for initially uncorrelated states in the limit $N\to\infty$~\cite{Carollo2021, Carollo2022, Fiorelli2023, Iemini2018, Piccito2021, Mukherjee2024}.

Applying this approximation to the exact equations of motion yields
\begin{equation}
\begin{aligned}
    \frac{d}{dt}m_x &= \kappa m_x m_z  - \frac{\Gamma}{2N} m_x, \\ \frac{d}{dt}m_y &= -\Omega m_z + \kappa m_y m_z  - \frac{\Gamma}{2N} m_y, \\ \frac{d}{dt}m_z &= \Omega m_y - \kappa \left(m_x^2 + m_y^2\right)  - \frac{\Gamma}{N} m_z,
\end{aligned}
\end{equation}
where $\kappa = \Gamma / 2$.

In the strict thermodynamic limit $N\to\infty$, the subleading terms vanish and the equations reduce to
\begin{equation}
\begin{split}
    \frac{d}{dt}m_x &= \kappa m_x m_z, \\
    \frac{d}{dt}m_y &= -\Omega m_z + \kappa m_y m_z, \\
    \frac{d}{dt}m_z &= \Omega m_y - \kappa \left(m_x^2 + m_y^2\right).
\end{split}
\end{equation}

These nonlinear equations exhibit persistent oscillations for $\kappa/\Omega < 1$~\cite{Iemini2018} (or in our original notation $\Gamma < 2\Omega$), corresponding to the BTC phase, and relaxational dynamics for $\kappa/\Omega > 1$.

\subsection{Failure for Correlated Initial States}

The mean-field approximation crucially relies on the assumption of weak correlations. It therefore fails for initially correlated states~\cite{Mukherjee2024}.

A particularly important example is the maximally mixed state in the symmetric subspace,
\begin{equation}
\rho(0) = \frac{\mathbf{1}}{d},
\end{equation}
where $d=2j+1=N+1$ is the dimension of the Hilbert space.

For this state, all first moments vanish,
\begin{equation}
m_x(0)=m_y(0)=m_z(0)=0,
\end{equation}
since $\Tr(J_\alpha)=0$.

Evaluating the mean-field equations at $t=0$ gives
\begin{equation}
\frac{d}{dt}\mathbf m(0)=\mathbf 0,
\end{equation}
so that
\begin{equation}
\mathbf m(t)=\mathbf 0
\quad \forall t.
\end{equation}

Thus, mean-field theory predicts trivial dynamics for this state. In contrast, the exact equations of motion yield a different result as we show below.

\section{Beyond Mean-Field Solution for a Maximally Mixed State}
\label{sec:beyond_mf_sol_mixed_state}
\subsection{Exact Equations and Initial Conditions}
We may obtain the exact equations of motion directly from the expressions derived in Appendix~\ref{sec:exact_eom} as
\begin{equation}
\begin{aligned}
    \frac{d}{dt}m_x(t) &= \frac{\Gamma}{N^2} \Bigl \langle J_x J_z + J_z J_x \Bigr \rangle(t) \\ &\quad -\frac{\Gamma}{2N} m_x(t), \\ \frac{d}{dt}m_y(t) &= -\Omega m_z(t)  \\ &\quad + \frac{\Gamma}{N^2} \Bigl \langle J_y J_z + J_z J_y \Bigr \rangle(t) \\ &\quad -\frac{\Gamma}{2N} m_y(t), \\ \frac{d}{dt}m_z(t) &= \Omega m_y(t)  \\ &\quad - \frac{2\Gamma}{N^2} \Bigl \langle J_x^2 + J_y^2 \Bigr \rangle(t) \\ &\quad -\frac{\Gamma}{N} m_z(t), 
\end{aligned}
\end{equation}
For the maximally mixed state $\rho(0) = \tfrac{1}{N+1} \mathbf 1$ all first moments vanish i.e. $\langle J_\alpha (0)\rangle =0$ for $\alpha=x,y,z$. However, $\langle J_x^2+J_y^2\rangle_{t=0} = \langle J^2 - J_z^2\rangle_{t=0} \neq 0$ and gives a finite contribution such that
\begin{equation}
\begin{aligned}
    \langle J^2-J_z^2\rangle_{t=0} &= j(j+1) - \Tr \Bigl (J_z^2 \, \rho(0)\Bigr) \\ & = \frac{N(N+2)}{4} - \frac{1}{N+1} \Tr \Bigl (J_z^2\Bigr)\\ & = \frac{N(N+2)}{4} - \frac{1}{N+1} \sum_{m_j=-j}^{+j} m_j^2,
\end{aligned}
\end{equation}
where we expressed $\Tr(J_z^2)$ in the $\ket{j,\, m_j}$ eigenbasis such that $J_z \ket{j,\, m_j} = m_j \ket{j,\, m_j}$ and $m_j=-j, -j+1, \ldots, +j$. Since the sum is over a symmetric range of an even function, one obtains
\begin{equation}
\begin{aligned}
    \sum_{m_j=-j}^{+j} m_j^2
    &= 2\sum_{m_j=1}^{+j} m_j^2 \\
    &= \frac{j(j+1)(2j+1)}{3}.
\end{aligned}
\end{equation}
Therefore, the full contribution reduces to
\begin{equation}
\begin{aligned}
    \langle J^2-J_z^2\rangle_{t=0} &= \frac{N(N+2)}{4}- \frac{N(N+1)(N+2)}{12(N+1)} \\ & = \frac{N(N+2)}{6}.
\end{aligned}
\end{equation}
In addition, mixed contributions vanish e.g. $\langle J_y J_z \rangle_{t=0} \propto \Tr(J_y J_z)=0$, which can be shown by expressing these in terms of spherical tensor operators and applying their orthogonality~\cite{Blum2012}. Therefore, at $t=0$ one obtains
\begin{equation}
    \frac{d}{dt}m_x(0) =0, \quad \frac{d}{dt}m_y(0) =0,
\end{equation}
while the $z$-component has a finite contribution
\begin{equation}
\begin{aligned}
    \frac{d}{dt}m_z(0)
    &= - \frac{2\Gamma}{N^2} \Bigl \langle J_x^2 + J_y^2 \Bigr \rangle_{t=0} \\
    & =-\frac{\Gamma(N+2)}{3N} \\
    &\xrightarrow[N \to \infty]{} -\frac{\Gamma}{3},
\end{aligned}
\end{equation}
implying that the beyond mean-field terms generate a non-zero initial time derivative of $m_z$.

%Thus, we have shown that the mean-field solution yields the incorrect dynamics for an initially maximally mixed state, highlighting that a) the mean-field description yields  initial state dependent behavior and b) there must be another mechanism generating initial state independent dynamics in the fully quantum description.

\subsection{Time-Dependent Solutions}
Having established the initial conditions, we now derive the time-dependent solutions. The system is governed by two competing timescales: coherent rotation set by $\Omega$ and dissipative processes set by $\Gamma$. In the BTC regime $\Gamma \ll \Omega$, allowing for a controlled expansion to leading order in $\Gamma$.

We first consider the mixed correlators such as $\langle J_y J_z\rangle$. For the maximally mixed initial state, these vanish at $t=0$. Any non-zero value must therefore be generated dynamically. Since this generation is itself induced by dissipation, it is $O(\Gamma)$. Combined with the explicit $\Gamma$ prefactor in the equations of motion, these contributions enter only at $O(\Gamma^2)$ and can be neglected to leading order.

In contrast, the term $\langle J_x^2 + J_y^2 \rangle$ is already non-zero at $t=0$. Its contribution is therefore $O(\Gamma)$ and must be retained. To leading order in $\Gamma$, the equations of motion reduce to
\begin{equation}
\begin{aligned}
    \mathbf{A}) \quad \frac{d}{dt}m_x(t) &= -\frac{\Gamma}{2N} m_x(t), \\ \mathbf{B}) \quad \frac{d}{dt}m_y(t) &= -\Omega m_z(t)  -\frac{\Gamma}{2N} m_y(t), \\ \mathbf{C}) \quad \frac{d}{dt}m_z(t) &= \Omega m_y(t)   -\frac{\Gamma}{N} m_z(t)\\ &\quad - \frac{2\Gamma}{N^2} \Bigl \langle J_x^2 + J_y^2 \Bigr \rangle_{t=0}.
\end{aligned}
\label{eq:first_order_coupled_odes}
\end{equation}
This separation of contributions highlights that to leading order in $\Gamma$, the dynamics of magnetization operators is governed by damped harmonic motion in the $(m_y, m_z)$ plane, supplemented by a constant source term.

Next, we derive a differential equation solely in terms of $m_z$. We begin by defining the constant
\begin{equation}
\begin{aligned}
    F_0:=\frac{2\Gamma}{N^2} \langle J_x^2 + J_y^2\rangle_{t=0}
    &= \frac{\Gamma(N+2)}{3N}
\end{aligned}
\end{equation}
Taking the time derivative of Eq.~(\ref{eq:first_order_coupled_odes}C)), and substituting Eq.~(\ref{eq:first_order_coupled_odes}B)) into the resulting expression yields
\begin{equation}
\begin{aligned}
    \frac{d^2}{dt^2}m_z(t)
    &= -\Omega^2 m_z(t)-\frac{\Omega\Gamma}{2N}m_y(t) \\
    &\quad -\frac{\Gamma}{N}\frac{d}{dt}m_z(t).
\end{aligned}
\label{eq:second_order_eq_almost_there}
\end{equation}
This equation still depends on $m_y(t)$. We therefore rearrange Eq.~(\ref{eq:first_order_coupled_odes}C)) to express $m_y(t)$ in terms of $m_z(t)$ and substitute it into Eq.~(\ref{eq:second_order_eq_almost_there}). This yields
\begin{equation}
\begin{gathered}
    \frac{d^2}{dt^2}m_z(t)
    +\Bigl(\frac{3\Gamma}{2N}\Bigr)\frac{d}{dt}m_z(t) \\
    {}+\Bigl(\Omega^2+\frac{\Gamma^2}{2N^2}\Bigr)m_z(t)
    =-\frac{\Gamma}{2N}F_0.
\end{gathered}
\end{equation}

This is the equation of a forced, damped harmonic oscillator with damping rate $\gamma:=\tfrac{3\Gamma}{2N}$, natural frequency $\omega_0^2:= \Omega^2 + \tfrac{\Gamma^2}{2N^2}$ and constant forcing related to $F:=-\tfrac{\Gamma F_0}{2N}$. 

%We emphasize that although taking derivatives and eliminating variables can make the driving term appear at higher order ($O\sim(\tfrac{\Gamma F_0}{N})$), the physically relevant scaling is determined from the original first-order differential equations (cf. Eq.~\ref{eq:first_order_coupled_odes}), where the source term enters at first order in $\Gamma$.

The general solution is obtained as the sum of the homogeneous solution and a particular integral, $m_z(t)=m_z^{h}(t)+m_z^{p}$. The homogenous solution is of the form
\begin{equation}
    m_z^h(t) = e^{-\tfrac{\gamma}2{}t} \Bigl(A \cos (\omega t) + B \sin(\omega t)\Bigr),
\end{equation}
where $\omega^2:= \omega_0^2 -(\tfrac{\gamma}{2})^2$ and $\gamma \ll \omega_0$. The particular solution is given by the constant
\begin{equation}
    m_z^{p} =\frac{F}{\omega_0^2}= -\frac{\Gamma N F_0}{2N^2 \Omega^2 + \Gamma^2}.
\end{equation}

Applying the initial conditions
\begin{equation}
    m_z(0)=0, \qquad \frac{d}{dt}m_z(0)=-F_0,
\end{equation}
we obtain
\begin{equation}
\begin{split}
    m_z(t)
   &  = m_z^p \Bigl(1 - e^{-\tfrac{\gamma}{2} t} \cos(\omega t)\Bigr) \\
    & \; -\frac{(F_0 +\tfrac{\gamma}{2} m_z^p)}{\omega} e^{-\tfrac{\gamma}{2} t} \sin(\omega t),
\end{split}
\label{eq:mixed_state_full_sol}
\end{equation}

The different terms contribute on distinct scales. The dominant contribution has prefactor $\tfrac{F_0}{\omega}$, while the correction proportional to $m_z^p$ scales as $O(\tfrac{\Gamma F_0}{N})$, and the most suppressed term scales as $O(\tfrac{\Gamma^2 F_0}{N^2 \omega})$. In the large-$N$ limit, the subleading contributions vanish, and the dynamics is well approximated by
\begin{equation}
    m_z(t)\approx - \frac{F_0}{\omega} e^{-\tfrac{\gamma}{2}t} \sin(\omega t).
\end{equation}
Using $F_0 \to \tfrac{\Gamma}{3}$ for $N\to\infty$, this reduces to
\begin{equation}
    m_z(t)\approx - \frac{\Gamma}{3\Omega} e^{-\tfrac{\gamma}{2}t} \sin(\omega t).
\end{equation}
In the strict thermodynamic limit $N\to \infty$, the decay rate vanishes and the dynamics reduces to persistent oscillations.

%Alternatively, one may take the thermodynamic limit directly in the first-order equations. In this case, the damping terms vanish while the source term $F_0$ remains finite. Although the source term disappears from the second-order equation upon differentiation, it reappears through the initial condition $\tfrac{d}{dt}m_z(0)=-F_0$, demonstrating that it continues to control the amplitude of the dynamics.

\begin{figure*}[tp]
    \centering
    \includegraphics[width=0.85\linewidth]{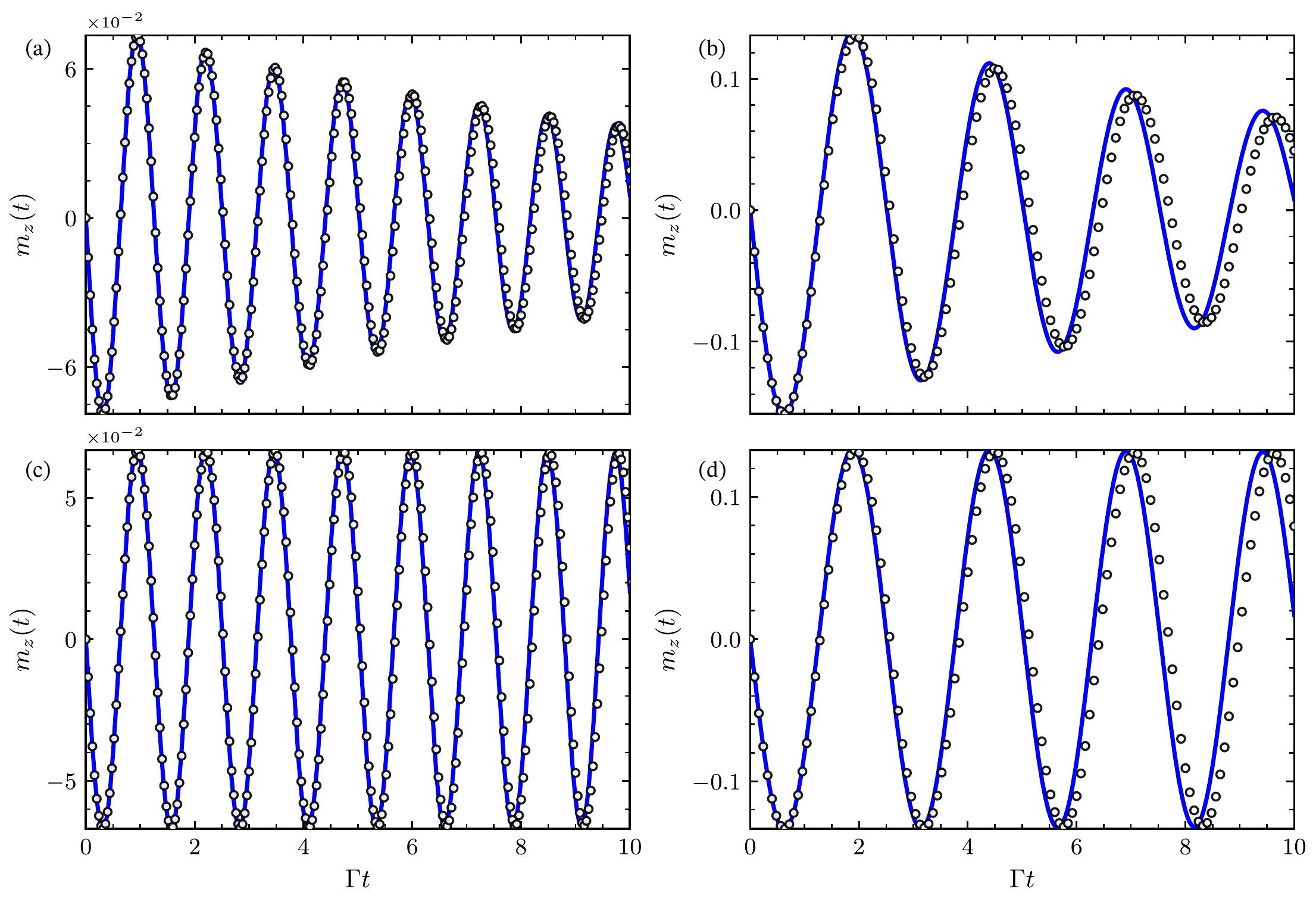}
    \caption{Magnetization Dynamics $m_z(t)$ for a Maximally Mixed Initial State. Solid blue lines denote the analytical predictions obtained deep in the BTC phase~[c.f. Eq.~\ref{eq:mixed_state_full_sol}], while white, hollow markers correspond to data sampled from numerically exact master equation solutions. In the first row, the system size is fixed to $N=10$ with columns corresponding to $\tilde{\Gamma}\equiv\Gamma/\Omega=0.2$ and $\tilde \Gamma =0.4$ respectively. The second row shows the corresponding results for system size $N=400$.}
    \label{fig:btc_mixed_state_dynamics}
\end{figure*}

We compare analytical solution presented in Eq.~(\ref{eq:mixed_state_full_sol}) to numerically exact master equation solutions obtained via the QuTiP package~\cite{Lambert2024} in Fig.~\ref{fig:btc_mixed_state_dynamics}. 

For weaker dissipation, corresponding to the regime closer to the extreme BTC limit, the analytical predictions exhibit excellent agreement with the exact numerical dynamics. As the dissipation strength is increased, deviations become more pronounced, particularly for smaller system sizes, consistent with the perturbative expansion in $\Gamma/N$. Furthermore, the oscillation amplitudes approach the predicted saturation value of $\Gamma/3$, in agreement with the analytical results.

These results demonstrate that dissipation does not merely damp dynamics, but can actively generate it through a non-unital source, motivating the need for a framework beyond mean-field theory.

\section{Liouvillian Eigenmode Treatment}
\label{sec:liouvillian_eigenmode_treatment}
\subsection{Non-Unital Dynamical Sourcing}
We now demonstrate how the source term emerges within a Liouvillian eigenmode description. To this end, we begin by noting that any density operator admits the generalized Bloch decomposition~\cite{Byrd2003, Kimura2003, Bertlmann2008, Byrd2011}
\begin{equation}
    \rho(t) = \frac{\mathbf 1}{d} + \delta\rho(t),
\end{equation}
where $\mathbf{1}$ is the identity operator, $d$ is the Hilbert space dimension and $\delta\rho(t)$ is the traceless component of $\rho(t)$, satisfying $\Tr[\delta\rho(t)]=0$.

Substituting this decomposition into the master equation, $\tfrac{d}{dt}\rho(t) = \mathcal{L}[\rho(t)]$, yields
\begin{equation}
\begin{aligned}
    \frac{d}{dt}\delta\rho(t) &= \mathcal{L}\big[\delta\rho(t)\big] + s, \\ s &:= \frac{1}{d}\mathcal{L}[\mathbf{1}].
\end{aligned}
\label{eq:traceless_master_eq}
\end{equation}
The second term acts as a constant source term in operator space. It vanishes for unital dynamics, but is non-zero whenever the Liouvillian is non-unital, i.e.,
\begin{equation}
    \mathcal{L}[\mathbf{1}] \neq 0.
\end{equation}

Since $\delta\rho(t)$ is traceless, it lies in the subspace orthogonal to the identity with respect to the Hilbert--Schmidt inner product. We may therefore expand it in a basis of right eigenoperators of $\mathcal{L}$ restricted to this traceless sector,
\begin{equation}
    \delta \rho(t) = \sum_\alpha c_\alpha(t) \, R_\alpha,
    \label{eq:traceless_rho_expansion}
\end{equation}
where $R_\alpha$ are traceless eigenoperators satisfying $\mathcal{L}[R_\alpha] = \lambda_\alpha R_\alpha$.

Physically, the fixed identity component acts as a constant inhomogeneous source for the traceless sector. This provides a mechanism for generating non-trivial dynamics even when the initial state has no mean-field polarization and constitutes the operator space origin of the inhomogeneous dynamics discussed above. 

To make this explicit, we follow the derivation presented in~\cite{Nemeth2026} to obtain the time-dependent form of $c_\alpha(t)$. Substituting Eq.~(\ref{eq:traceless_rho_expansion}) into Eq.~(\ref{eq:traceless_master_eq}) yields
\begin{equation}
    \sum_\alpha \dot{c}_\alpha(t)\, R_\alpha
    =
    \sum_\alpha c_\alpha(t)\, \mathcal L[R_\alpha] + s.
\end{equation}
Using $\mathcal{L}[R_\alpha] = \lambda_\alpha R_\alpha$, multiplying from the left by $L_\beta^\dagger$, and applying the biorthonormality condition $\Tr(L_\beta^\dagger R_\alpha) = \delta_{\alpha\beta}$, we obtain
\begin{equation}
    \dot{c}_\alpha(t) - \lambda_\alpha c_\alpha(t) = s_\alpha,
\end{equation}
where $s_\alpha := \Tr(L_\alpha^\dagger s)$, and we have relabelled $\beta \leftrightarrow \alpha$.

This first-order differential equation can be solved using an integrating factor, giving
\begin{equation}
    c_\alpha(t)
    =
    e^{\lambda_\alpha t} c_\alpha(0)
    +
    \frac{e^{\lambda_\alpha t}-1}{\lambda_\alpha}\, s_\alpha,
\end{equation}
valid for $\lambda_\alpha \neq 0$. 
The first term corresponds to the usual initial state dependent evolution of the eigenmode, while the second term represents an initial state independent contribution arising from the source term.

To evaluate $s_\alpha$, we temporarily switch to the vectorized formalism~\cite{Gyamfi2020}, where the eigenvalue equations take the form
\begin{equation}
\begin{aligned}
    \mathbb L \ket{R_\alpha}\rangle &= \lambda_\alpha \ket{R_\alpha}\rangle, \\ \mathbb L^\dagger \ket{L_\alpha}\rangle &= \lambda_\alpha^* \ket{L_\alpha}\rangle.
\end{aligned}
\label{eq:vectorized_liouville_eigenvalue_eqs}
\end{equation}
The inner product is defined as
\begin{equation}
    \langle\!\langle A \vert B \rangle\!\rangle := \Tr(A^\dagger B),
\end{equation}
so that the adjoint of $\mathbb S$ satisfies
\begin{equation}
    \langle\!\langle A \vert \mathbb S B \rangle\!\rangle
    =
    \langle\!\langle \mathbb S^\dagger A \vert B \rangle\!\rangle.
\end{equation}

Using this, we can write
\begin{equation}
\begin{aligned}
    \Tr(L_\alpha^\dagger \mathcal L[\mathbf{1}])
    &= \langle\!\langle L_\alpha \vert \mathbb L\, \mathbf{1} \rangle\!\rangle \\
    &= \langle\!\langle \mathbb L^\dagger L_\alpha \vert \mathbf{1} \rangle\!\rangle.
\end{aligned}
\end{equation}
Applying Eq.~(\ref{eq:vectorized_liouville_eigenvalue_eqs}) then gives
\begin{equation}
    \begin{split}
        \langle\!\langle \mathbb L^\dagger L_\alpha \vert \mathbf{1} \rangle\!\rangle
        &= \langle\!\langle \lambda_\alpha^* L_\alpha \vert \mathbf{1} \rangle\!\rangle \\
        &= \lambda_\alpha \langle\!\langle L_\alpha \vert \mathbf{1} \rangle\!\rangle \\
        &= \lambda_\alpha \Tr(L_\alpha^\dagger \mathbf{1}).
    \end{split}
\end{equation}

Therefore,
\begin{equation}
\begin{aligned}
    s_\alpha
    &= \frac{1}{d}\Tr(L_\alpha^\dagger \mathcal L[\mathbf{1}]) \\
    &= \frac{\lambda_\alpha}{d}\Tr(L_\alpha^\dagger \mathbf 1).
\end{aligned}
\end{equation}

This shows that the initial state independent sourcing of a given eigenmode $\alpha$ is controlled by the overlap of the corresponding left eigenoperator with the identity. In particular, only those modes with $\Tr(L_\alpha^\dagger \mathbf 1)\neq 0$ receive a finite contribution from the source term.

Finally, the solution $\delta \rho(t)$ becomes
\begin{equation}
\begin{aligned}
    \delta\rho(t) &= \sum_\alpha e^{\lambda_\alpha t} \, \mathrm{Tr}\!\big(L^\dagger_\alpha \, \delta\rho(0)\big)\, R_\alpha \\ &\quad + \sum_\alpha \frac{e^{\lambda_\alpha t}-1}{d} \, \mathrm{Tr}\!\big(L^\dagger_\alpha \mathbf 1\big)\, R_\alpha.
\end{aligned}
\end{equation}

For the maximally mixed initial state,
\begin{equation}
\rho(0) = \frac{\mathbf{1}}{d},
\qquad
\delta\rho(0)=0,
\end{equation}
the homogeneous contribution vanishes, yielding
\begin{equation}
\delta\rho(t)
=
\sum_\alpha
\frac{e^{\lambda_\alpha t}-1}{d}
\, \mathrm{Tr}\!\big(L_\alpha^\dagger \mathbf 1 \big)\, R_\alpha.
\end{equation}
Therefore, the initial state independent, sourced contribution to the evolution of some observable is given by
\begin{equation}
\begin{aligned}
    \langle O(t) \rangle\eval_{\mathrm{s}}= \sum_\alpha \frac{e^{\lambda_\alpha t}-1}{d} \, \Tr\!\big(L_\alpha^\dagger \mathbf 1 \big)\, \Tr \!\big(O R_\alpha \big).
\end{aligned}
\end{equation}

Before discussing the physical consequences of this contribution, it is important to verify that it survives in the intensive magnetisation,
\begin{equation}
    m_z(t)=\frac{2}{N}\langle J_z(t)\rangle ,
\end{equation}
in the thermodynamic limit. We therefore estimate the scaling of each factor separately. The dimension of the collective spin Hilbert space is
\begin{equation}
    d=2j+1=N+1,
\end{equation}
so that the prefactor scales as
\begin{equation}
    \frac{1}{d}=\frac{1}{N+1}\sim O(N^{-1}).
\end{equation}

The overlap of the left eigenoperator with the identity can be written in terms of the normalized rank-zero spherical tensor
\begin{equation}
    T^0_0=\frac{\mathbf 1}{\sqrt{N+1}}.
\end{equation}
Hence
\begin{equation}
    \Tr(L_\alpha^\dagger \mathbf 1)
    =
    \sqrt{N+1}\,
    \Tr(L_\alpha^\dagger T^0_0).
\end{equation}
Expanding the left eigenoperator in the Hilbert--Schmidt orthonormal spherical tensor basis,
\begin{equation}
    L_\alpha
    =
    l^{(\alpha)}_{0,0} T^0_0
    +
    \sum_{(k,q)\neq(0,0)}
    l^{(\alpha)}_{k,q} T^k_q ,
\end{equation}
with
\begin{equation}
    \Tr\!\left[(T^{k'}_{q'})^\dagger T^k_q\right]
    =
    \delta_{kk'}\delta_{qq'},
\end{equation}
we obtain
\begin{equation}
    \Tr(L_\alpha^\dagger T^0_0)
    =
    \left(l^{(\alpha)}_{0,0}\right)^* .
\end{equation}
Therefore
\begin{equation}
    \Tr(L_\alpha^\dagger \mathbf 1)
    \sim
    N^{1/2} \left(l^{(\alpha)}_{0,0}\right)^* .
\end{equation}
This contribution is finite and non-vanishing after normalisation provided the relevant left eigenoperator retains an $O(1)$ projection onto the identity sector, i.e.
\begin{equation}
    l^{(\alpha)}_{0,0}=O(1).
\end{equation}

Next, consider the overlap of the right eigenoperator with $J_z$. Since $J_z$ is proportional to the rank-one spherical tensor $T^1_0$, we may write
\begin{equation}
    J_z
    =
    \sqrt{\Tr(J_z^2)}\,T^1_0 ,
\end{equation}
up to a convention-dependent sign. For the collective spin sector $j=N/2$,
\begin{equation}
\begin{aligned}
    \Tr(J_z^2)
    &= \frac{1}{3}j(j+1)(2j+1) \\
    &= \frac{N(N+1)(N+2)}{12} \sim O(N^3),
\end{aligned}
\end{equation}
and therefore
\begin{equation}
    \sqrt{\Tr(J_z^2)}\sim O(N^{3/2}).
\end{equation}
Expanding the right eigenoperator as
\begin{equation}
    R_\alpha
    =
    r^{(\alpha)}_{1,0} T^1_0
    +
    \sum_{(k,q)\neq(1,0)}
    r^{(\alpha)}_{k,q} T^k_q ,
\end{equation}
we find
\begin{equation}
\begin{aligned}
    \Tr(J_z R_\alpha)
    &= \sqrt{\Tr(J_z^2)}\, \Tr(T^1_0 R_\alpha) \sim N^{3/2} r^{(\alpha)}_{1,0}.
\end{aligned}
\end{equation}
Thus this overlap is non-vanishing in the thermodynamic limit provided the relevant right eigenoperator has an $O(1)$ projection onto the rank-one sector,
\begin{equation}
    r^{(\alpha)}_{1,0}=O(1).
\end{equation}

Combining the scalings, the contribution to the intensive magnetisation behaves as
\begin{equation}
\begin{aligned}
   &= \frac{2}{N} \frac{1}{N+1} \Tr(L_\alpha^\dagger \mathbf 1) \Tr(J_z R_\alpha)  \\ &\sim N^{-1} N^{-1} N^{1/2} N^{3/2}   = O(1).
\end{aligned}
\end{equation}
Therefore this term gives a finite contribution to the intensive magnetisation in the thermodynamic limit, provided
\begin{equation}
    l^{(\alpha)}_{0,0}=O(1),
    \qquad
    r^{(\alpha)}_{1,0}=O(1).
\end{equation}
Physically, these conditions mean that the left eigenoperator has a finite overlap with the identity sector, while the right eigenoperator has a finite overlap with the observable sector probed by $J_z$.

Thus, the remaining question is whether these overlaps remain finite. In the main text, we argue that this is ensured by NH topology. In particular, a non-zero local winding number signals an obstruction to eigenmode localization in operator space. As a result, the relevant oscillatory modes cannot remain confined to a single sector, but instead acquire finite weight across neighbouring sectors. In the present case, this necessarily includes both the identity (source) sector and the rank-one sector probed by $J_z$, thereby ensuring non-vanishing overlaps.

\subsection{Higher-Order Spin Correlations}
Our previous argument naturally extends to higher-order spin correlators. Consider operators of the form
\begin{equation}
    O_k = c_k T^k_0,
    \label{eq:higher_spin_corr_tensor_k}
\end{equation}
where $O_k$ is a degree-$k$ polynomial in collective spin operators. For example, one has $T^2_0 \propto (3J_z^2 - J^2)$ and $T^3_0 \propto (5J_z^3 - 3J_z J^2)$ (up to normalization and convention-dependent corrections).

We now determine the coefficient $c_k$. Taking the inner product with $T^k_0$ yields
\begin{equation}
\begin{aligned}
    \Tr\!\left((T^k_0)^\dagger O_k\right)
    &= c_k \Tr\!\left((T^k_0)^\dagger T^k_0\right) \\
    &= c_k,
\end{aligned}
\end{equation}
where we used the orthonormality of spherical tensor operators. On the other hand,
\begin{equation}
    \Tr\!\left((T^k_0)^\dagger O_k\right)
    =
    \frac{1}{c_k^*} \Tr(O_k^\dagger O_k),
\end{equation}
which follows from Eq.~(\ref{eq:higher_spin_corr_tensor_k}). Combining these expressions gives
\begin{equation}
    \Tr(O_k^\dagger O_k) = \vert c_k \vert^2,
\end{equation}
and for Hermitian operators $O_k$, this reduces to $c_k = \sqrt{\Tr(O_k^2)}$ up to a phase.

We now estimate the scaling of $\Tr(O_k^2)$. Since $O_k$ is a degree-$k$ polynomial in collective spin operators, its matrix elements scale as $O_k \sim N^k$. Taking the trace over a Hilbert space of dimension $d=N+1$, one finds
\begin{equation}
    \Tr(O_k^2) \sim d \times N^{2k} \sim N^{2k+1},
\end{equation}
and therefore
\begin{equation}
    \sqrt{\Tr(O_k^2)} \sim N^{k+1/2}.
\end{equation}

Following the derivation in the previous section, we can now show that an intensive observable defined as
\begin{equation}
    \tilde{O}_k = \left(\frac{2}{N}\right)^k O_k
\end{equation}
has a finite thermodynamic sourced response. Indeed, the relevant contribution scales as
\begin{equation}
\begin{aligned}
    &=\left(\frac{2}{N}\right)^k \frac{1}{N+1} \Tr(L_\alpha^\dagger \mathbf{1}) \Tr(O_k R_\alpha) \\ &\sim N^{-k} N^{-1} N^{1/2} N^{k+1/2}  = O(1),
\end{aligned}
\end{equation}
where we used $\Tr(L_\alpha^\dagger \mathbf{1}) \sim N^{1/2}$ and $\Tr(O_k R_\alpha) \sim N^{k+1/2}$.

The resulting contribution is non-vanishing provided the overlaps
\begin{equation}
\begin{aligned}
    l^{(\alpha)}_{0,0} &= \Tr(L_\alpha^\dagger T^0_0), \\ r^{(\alpha)}_{k,0} &= \Tr\!\left((T^k_0)^\dagger R_\alpha\right)
\end{aligned}
\end{equation}
remain finite in the thermodynamic limit. In contrast to the magnetization case ($k=1$), this now requires the relevant eigenmodes to be delocalized over an operator space distance of order $k$. Thus, higher-order correlators provide a more stringent probe of operator space delocalization, as the contributing modes must extend further from the identity sector to the corresponding rank-$k$ sector.

\subsection{Constraints on Initial Conditions}
In this section, we highlight how non-unital contributions impose additional constraints on the initial dynamics of an observable. In particular, these contributions enter through the initial time derivative $\tfrac{d}{dt}\langle O(0) \rangle$. Using Eq.~(\ref{eq:traceless_master_eq}), we obtain
\begin{equation}
\begin{aligned}
    \frac{d}{dt}\langle O(0)\rangle = \Tr \Bigl( O \frac{d}{dt} \rho(0)\Bigr)  &= \Tr \Bigl( O \, \mathcal L [\delta \rho(0)]\Bigr) \\ &\quad + \frac{1}{d}\Tr \Bigl( O \, \mathcal L[\mathbf 1]\Bigr),
\end{aligned}
\end{equation}
which explicitly separates the traceless contribution from the non-unital term originating from the identity sector.

For a maximally mixed initial state, $\rho(0) = \tfrac{1}{d}\mathbf 1$, one has $\delta\rho(0)=0$, so that the first term vanishes. In this case, the initial dynamics is entirely determined by the non-unital contribution. Using $\mathcal{L} [\mathbf 1] = -\tfrac{2\Gamma}{N} J_z$, we obtain
\begin{equation}
    \frac{d}{dt}m_z(0)= -\frac{\Gamma(N+2)}{3N} \equiv -F_0,
\end{equation}
where $m_z = \tfrac{2}{N}\langle J_z\rangle$. This reproduces the same quantity $F_0$ obtained from the explicit beyond-mean-field calculation in Sec..~\ref{sec:beyond_mf_sol_mixed_state}.

\section{Hierarchy of Spin Correlations}
\label{sec:hierarchy_of_spin_correlations}

In this section, we further elaborate on the hierarchy of collective spin correlations arising from the $(2j+1)$ allowed tensor-rank sectors. As proven in Sec.~\ref{sec:liouvillian_eigenmode_treatment}, intensive higher-order spin correlations retain a finite sourced contribution in the thermodynamic limit. 

Using the effective hopping model derived in Sec.~\ref{sec:mapping_nh_hopping_model} [Eq.~(\ref{eq:vectorized_non_hermitian_hopping_model})], we observe that each hopping process along the $k$-coordinate is generated by the dissipative dynamics and therefore carries a factor of $\Gamma$. Consequently, deep within the BTC phase, for the non-unital source located at $k=0$ to generate a response in an intensive degree-$k$ observable, the dynamics must involve at least $k$ successive hopping processes, leading to an overall scaling proportional to $\Gamma^k$. We therefore obtain the scaling law
\begin{equation}
    \langle \tilde{O}_k \rangle \sim \Gamma^k,
\end{equation}
where $\tilde{O}_k$ denotes an intensive collective spin operator of degree $k$.

Importantly, this scaling relation applies specifically to the sourced contribution to the dynamics. Numerically, this response can be isolated from the initial state dependent component by initializing the system in the maximally mixed state, for which
\begin{equation}
    \delta \rho(0)
    =
    \rho(0)-\frac{\mathbf 1}{d}
    =
    0.
\end{equation}
In this case, the evolution of $\delta \rho(t)$ is purely driven by the initial state independent sourced dynamics.

\begin{figure*}[tp]
    \centering
    \includegraphics[width=0.98\linewidth]{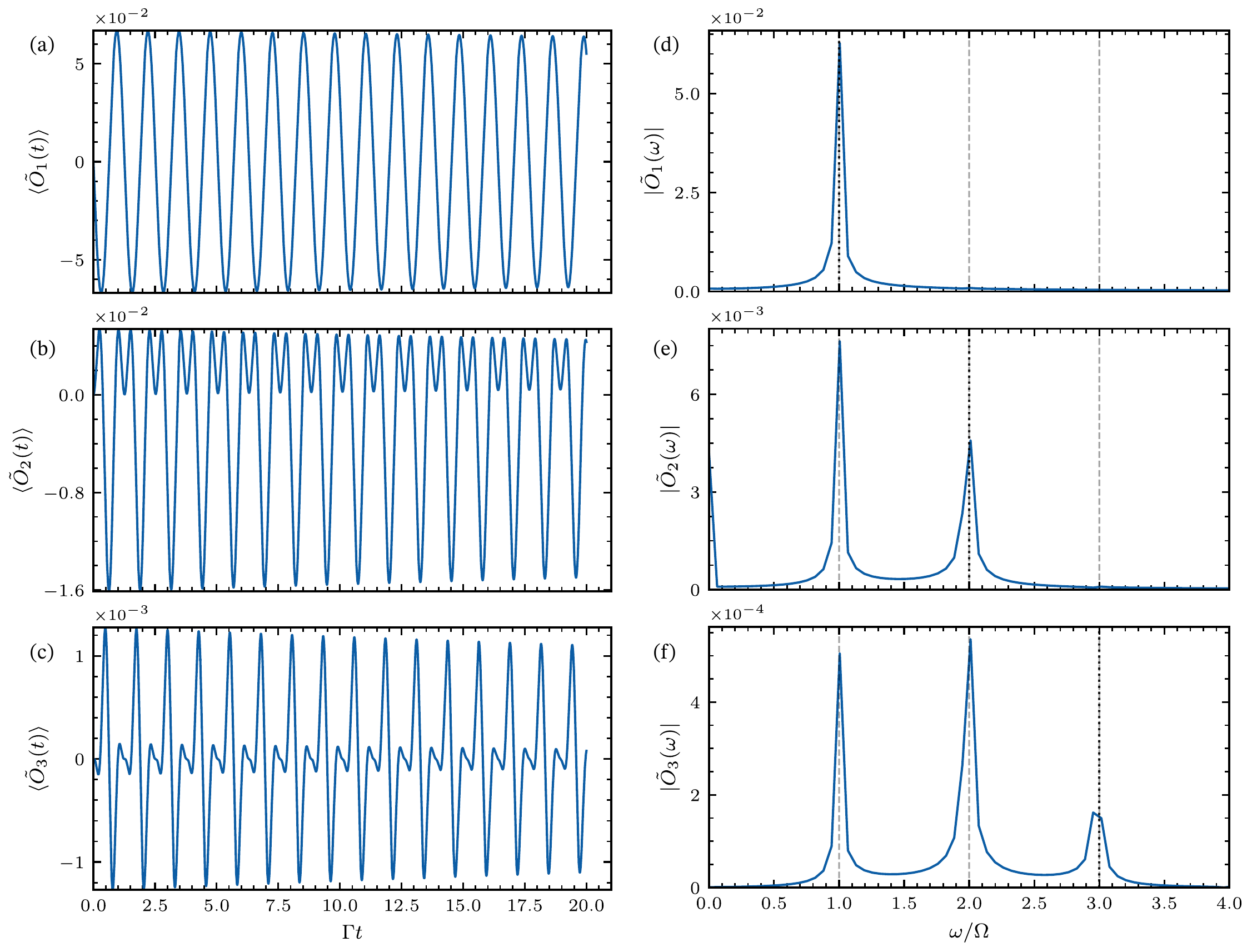}
    \caption{Time-series and frequency spectra for higher-order correlation functions. (a)-(c) show the time series of $\langle \tilde{O}_{1,2,3}(t)\rangle$, defined in Eq.~(\ref{eq:higher_spin_corr_funcs_defns}). The corresponding frequency spectra are shown in (d)-(f) respectively, where
    $\vert\tilde{O}_k(\omega)\vert = \left|\mathcal{FT}\!\left[\langle \tilde{O}_k(t)\rangle\right]\right\vert$
    denotes the Fourier transform of the expectation value of the degree-$k$ intensive observable. In all panels, $N=320$ and $\tilde{\Gamma}\equiv \Gamma/\Omega = 0.2$. The system is initialized in the maximally mixed state, $\rho(0)=\mathbf{1}/d$, in order to isolate the sourced contributions. Dashed lines indicate the positions of the $n\Omega$ harmonics for integer $n$.}
    \label{fig:higher_order_corrs_time_series}
\end{figure*}

We make our claims concrete by considering the three intensive operators
\begin{equation}
    \begin{split}
        & \tilde{O}_1 = \Bigl(\frac{1}{j}\Bigr) J_z, \\
        & \tilde{O}_2 = \Bigl(\frac{1}{j}\Bigr)^2 \Bigl[3J_z^2 - J^2\Bigr], \\
        & \tilde{O}_3 = \Bigl(\frac{1}{j}\Bigr)^3 \Bigl[5J_z^3 - 3 J_z J^2\Bigr],\end{split}
    \label{eq:higher_spin_corr_funcs_defns}
\end{equation}
which are proportional to $T^1_0, T^2_0$ and $T^3_0$ respectively. We showcase the time-series for each of these observables in panels a), b) and c) of Fig.~\ref{fig:higher_order_corrs_time_series}. In the weak dissipation limit, all three observables exhibit persistent oscillations, with behaviors getting progressively richer. As discussed in the main text, for $k=2$ there are two topological oscillatory pairs. In general, for a rank-$k$ tensor, there are $k$ distinct oscillatory frequencies, which is also illustrated in panels d), e) and f) of Fig.~\ref{fig:higher_order_corrs_time_series}.

\begin{figure*}[tp]
    \centering
    \includegraphics[width=0.98\linewidth]{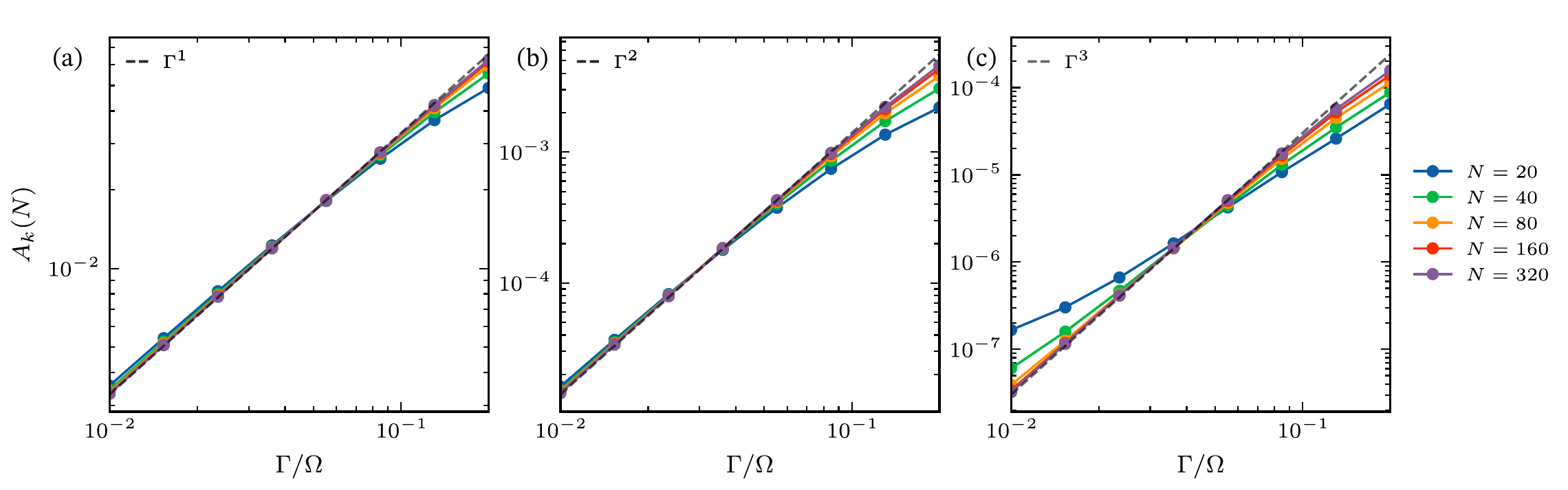}
    \caption{Finite-$N$ analysis of the $\Gamma^k$ scaling law. (a) shows the dependence of $A_k$ for $k=1$ on $\tilde{\Gamma} \equiv \Gamma/\Omega$ in the weak dissipation regime for $N\in\{20,40,80,160,320\}$. The same analysis is presented for $k=2$ and $k=3$ in  (b) and (c) respectively. A dashed line shows the predicted scaling of $\Gamma^k$.} 
    \label{fig:scaling_corrs}
\end{figure*}

\begin{figure*}[tp]
    \centering
    \includegraphics[width=0.95\linewidth]{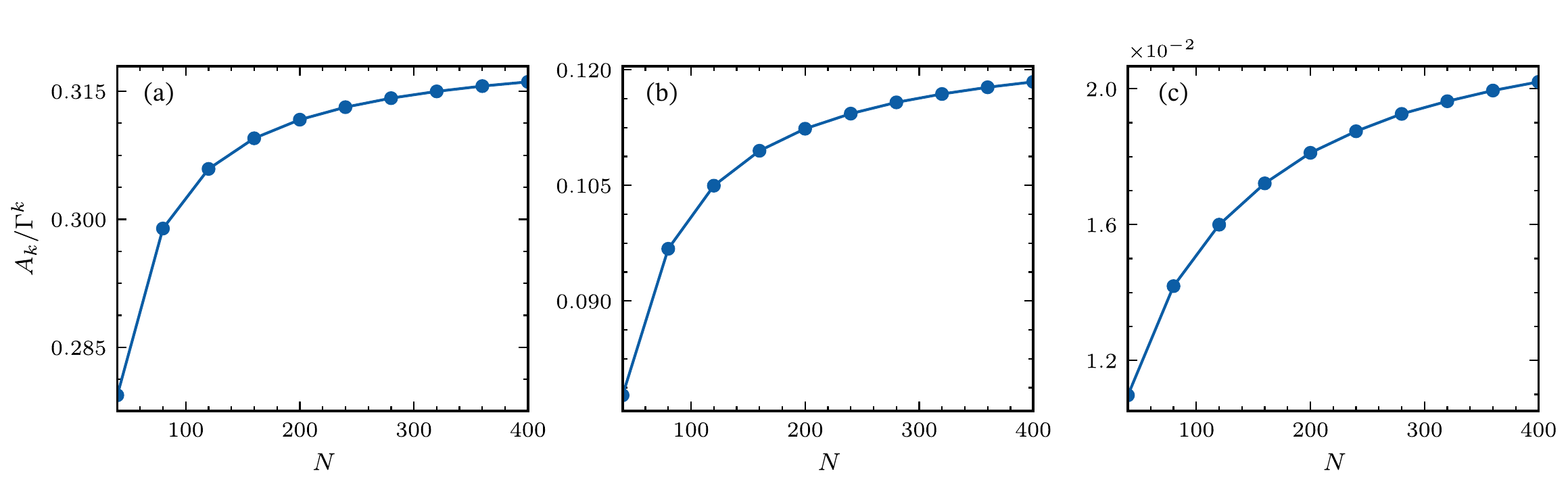}
    \caption{Saturation of $A_k /\Gamma^k$ for large-$N$. (a) shows the dependence of $A_k/\Gamma^k$ for $k=1$ on $N$ in the weak dissipation regime evaluated at $N\in\{40, 80, 120, 160, 200, 240, 280, 320, 360, 400\}$ and $\tilde{\Gamma} \equiv \Gamma/\Omega = 0.20$. The same analysis is presented for $k=2$ and $k=3$ in (b) and (c) respectively.}
    \label{fig:saturation_corrs}
\end{figure*}

Next, we demonstrate that the leading-order sourced contributions obey the scaling law $A_k \sim \Gamma^k$,
where $A_k$ denotes the Fourier amplitude of the relevant oscillatory mode associated with a degree-$k$ intensive observable. The specific modes analyzed are indicated by the dotted lines in panels d), e), and f) of Fig.~\ref{fig:higher_order_corrs_time_series}.

In the numerical analysis, the Fourier amplitude is extracted by targeting the expected harmonic frequency ($\Omega, 2\Omega$, or $3\Omega$) and subsequently performing a local search in frequency space to identify the corresponding spectral peak. 

We demonstrate this scaling explicitly in Fig.~\ref{fig:scaling_corrs}. In each case, the finite-$N$ numerical results show close agreement with the expected $\Gamma^k$ scaling behavior.

Furthermore, we demonstrate the emergence of a saturation trend in the large-$N$ limit in Fig.~\ref{fig:saturation_corrs}, where $A_k/\Gamma^k$ approaches a finite value. This is consistent with the scaling analysis presented earlier and confirms that the sourced contributions remain finite in the thermodynamic limit.

\section{Spectral Localizer}
\label{sec:spectral_localizer}
In this section, we provide additional details on the implementation of the spectral localizer, with particular emphasis on the choice of the parameter $\kappa$ (see Eq.~(\ref{eq:spectral_localizer}) of the main text).

In Ref.~\cite{Jezequel2025} it was shown that, in one dimension, the point-gap winding marker arises as the leading non-vanishing term in a controlled small-$\kappa$ expansion of the spectral localizer index. In practice, there exists a finite $\kappa$-interval, a stability window, within which the index faithfully captures the underlying point-gap topology \cite{Cerjan2024}. Throughout this work, we fix $\kappa = 1$, chosen within the numerically verified $\kappa$-stable regime, where the localizer gap remains open and the topological classification is stable under variations in $\kappa$. In this regime, the index faithfully captures the intrinsic point-gap topology while resolving its spatial structure. The parameter $\kappa$ controls the relative weighting between spectral and positional contributions in the localizer, with $\kappa^{-1}$ setting the effective spatial resolution scale over which the topology is probed \cite{Cerjan2024}.

\begin{figure*}[tp]
    \centering
    \includegraphics[width=0.75\linewidth]{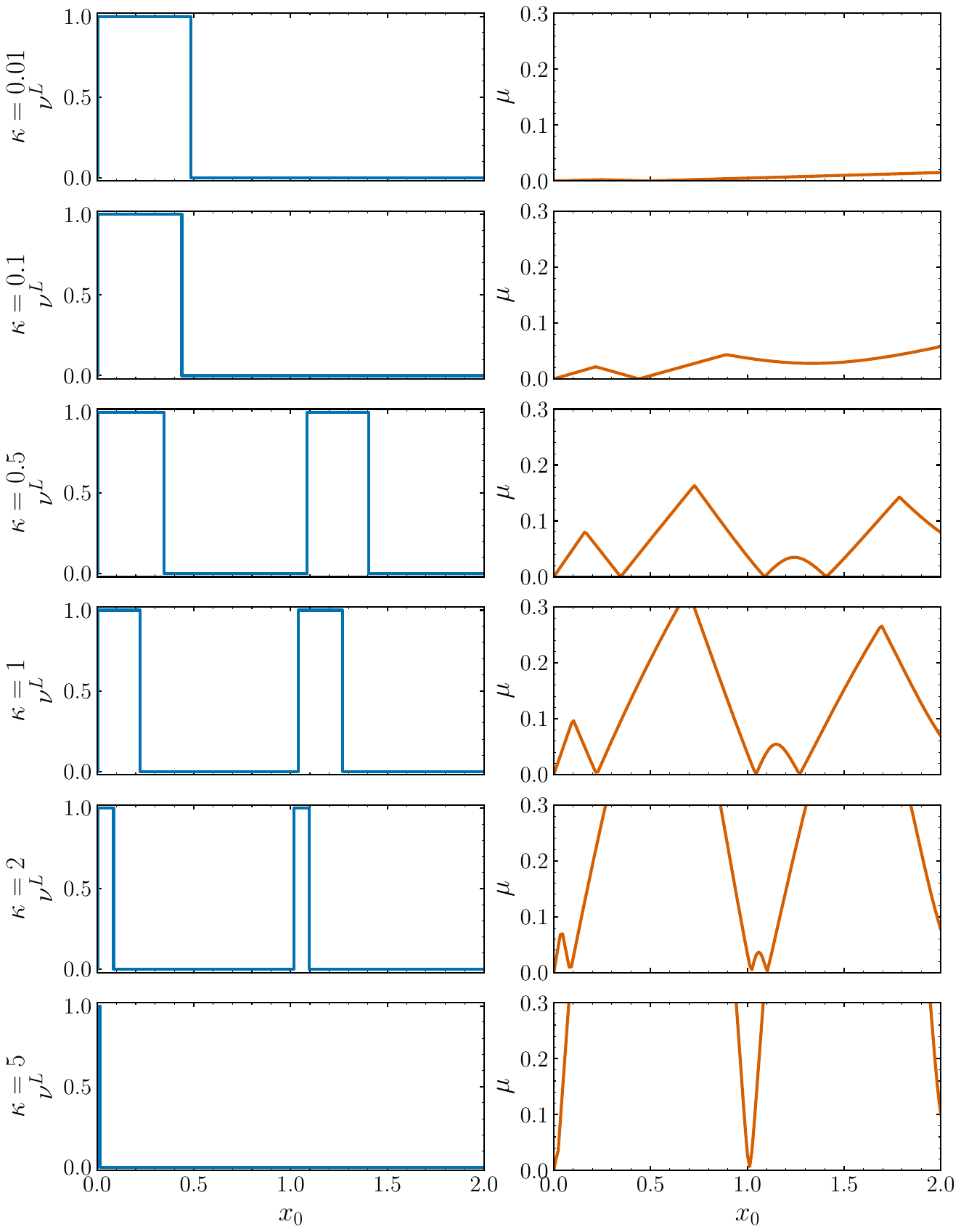}
    \caption{Topological domains as a function of $\kappa$. We show the dependence of the localizer index $\nu^L$ and gap $\mu$ on the resolution parameter $\kappa$ for $\kappa\in\{0.01,0.1,0.5,1,2,5\}$ and fixed $\lambda_0=0$. A stable $\kappa$-window is observed in which the gap remains open (inside topologically non-trivial regions) and the topological classification is unchanged. Results are shown for fixed $\tilde{\Gamma}\equiv \Gamma/\Omega = 1$, system size $N=10$ and are truncated to $x_0 \leq2$ (beyond which $\nu^L$ remains zero.).}
    \label{fig:sup1}
\end{figure*}

\begin{figure*}[tp]
    \centering
    \includegraphics[width=0.75\linewidth]{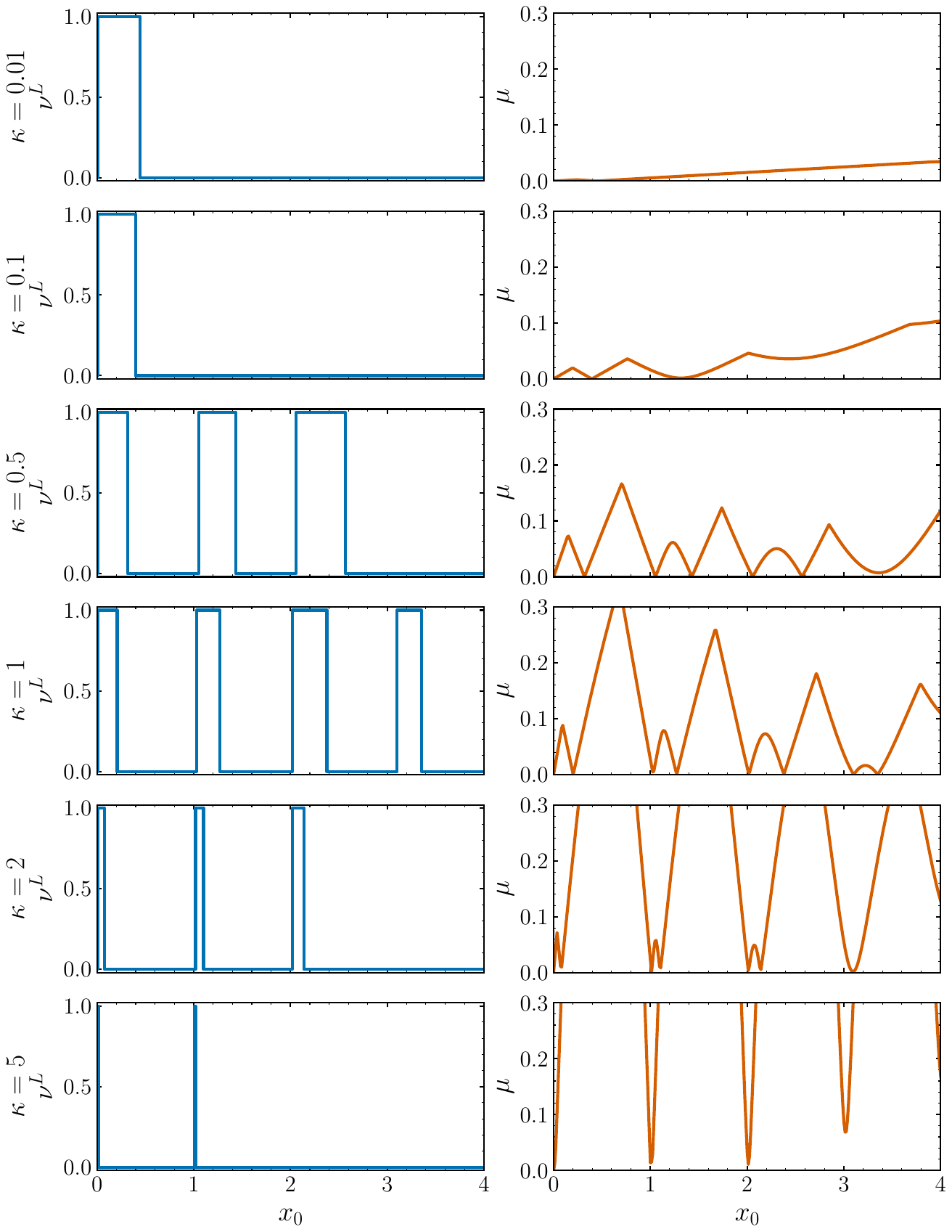}
    \caption{Topological domains as a function of $\kappa$. We show the dependence of the localizer index $\nu^L$ and gap $\mu$ on the resolution parameter $\kappa$ for $\kappa\in\{0.01,0.1,0.5,1,2,5\}$ and fixed $\lambda_0=0$. A stable $\kappa$-window is observed in which the gap remains open (inside topologically non-trivial regions) and the topological classification is unchanged. Results are shown for fixed $\tilde{\Gamma}\equiv \Gamma/\Omega = 1$, system size $N=20$ and are truncated to $x_0 \leq4$ (beyond which $\nu^L$ remains zero.).}
    \label{fig:sup2}
\end{figure*}

In Figs.~\ref{fig:sup1} and \ref{fig:sup2} we demonstrate that the non-trivial topology remains stable over a broad window of $\kappa$. For small $\kappa$, the localizer probes the system over an effectively large spatial scale, rendering it insensitive to fine, spatially resolved structure. As a result, smaller topological domains are smeared out. In the intermediate regime $\kappa \in [0.5,2]$, the localizer achieves optimal resolution, clearly resolving localized topological domains. For larger $\kappa$, however, the probe becomes overly localized and begins to over-resolve short-scale changes.

We further examine the behavior of the localizer gap $\mu_{(x_0,\lambda_0)}$ as a function of $x_0$. Within topologically non-trivial regions the gap remains open, while it exhibits sharp dips at the boundaries between domains, signaling local topological transitions.

\begin{figure*}[tp]
    \centering
    \includegraphics[width=0.75\linewidth]{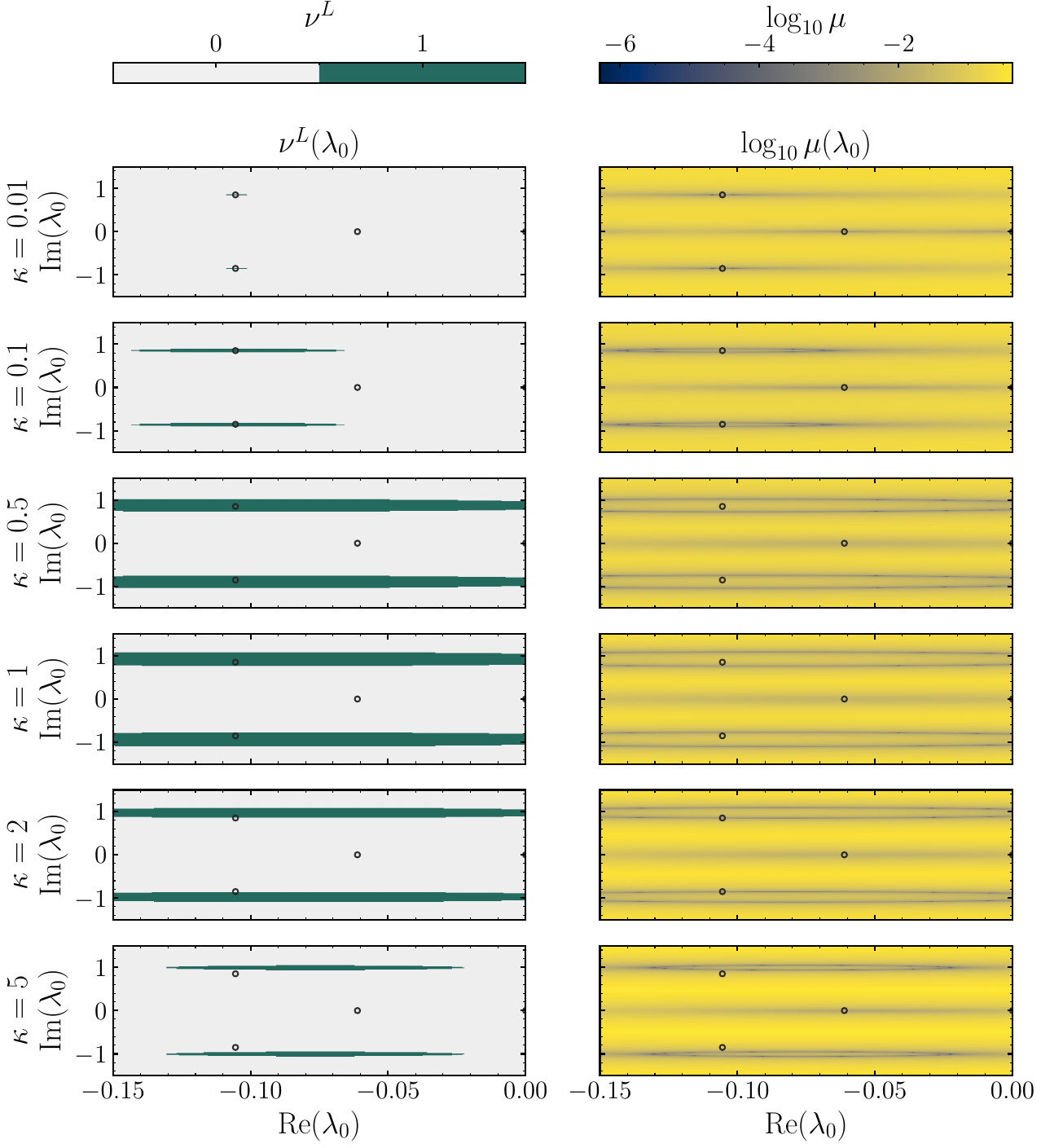}
    \caption{Spectral islands as a function of $\kappa$. We show the dependence of the localizer index $\nu^L$ and gap $\mu$ on the resolution parameter $\kappa$ for $\kappa\in\{0.01,0.1,0.5,1,2,5\}$ and fixed $x_0=1$. Hollow circular markers denote the eigenvalues of the Liouvillian $\mathcal L$. A stable $\kappa$-window is observed in which the gap remains open (inside topologically non-trivial regions) and the topological classification is unchanged. Results are shown for fixed $\tilde{\Gamma}\equiv \Gamma/\Omega = 1$ and system size $N=10$. Here, we focus on the slowest-decaying oscillatory pair of eigenmodes.}
    \label{fig:sup3}
\end{figure*}

In addition, we probe the robustness of the topology by sweeping $\lambda_0$ at fixed position $x_0$, as shown in Fig.~\ref{fig:sup3}. We focus on the slowest-decaying oscillatory eigenmodes, corresponding to eigenvalues with the least negative $\mathrm{Re}(\lambda)$, and fix $x_0 = 1$. The topological islands are most clearly resolved within the intermediate window $\kappa \in [0.5,2]$, where the localizer gap $\mu$ remains open throughout these regions.

\end{document}